\documentclass[12pt,preprint]{aastex}

\newcommand\msun{\ifmmode{\hbox{M$_\odot$}}\else{M$_\odot$}\fi}
\newcommand{\reduceme}{\mbox{R\raisebox{-0.35ex}{E}D%
\hspace{-0.05em}\raisebox{0.85ex}{uc}\hspace{-0.90em}%
\raisebox{-.35ex}{{m}}\hspace{0.05em}E}}

\slugcomment{}

\shorttitle{Spectroscopy of Globular Clusters in NGC\,1407}
\shortauthors{Cenarro et al.}

\begin{document}

\title{Stellar Populations of Globular Clusters in the Elliptical Galaxy NGC\,1407}
\author{A. Javier Cenarro \altaffilmark{1,2} and Michael A. Beasley \altaffilmark{2}} 
\affil{Instituto de Astrof\'{\i}sica de Canarias, La Laguna, 38200, Canary Islands, Spain} 
\email{cenarro@iac.es}  

\author{Jay Strader and Jean P. Brodie} 
\affil{Lick Observatory, University of California, CA 95064, USA} 
\and
\author{Duncan A. Forbes} 
\affil{Centre for Astrophysics \& Supercomputing, Swinburne University, Hawthorn VIC 3122, Australia}
\altaffiltext{1}{Also at Departamento de F\'{\i}sica de la Tierra, Astronom\'{\i}a y Astrof\'{\i}sica II, Facultad de F\'{\i}sicas, Universidad Complutense de Madrid, 28040 Madrid, Spain}
\altaffiltext{2}{Also at Lick Observatory, University of California, CA 95064, USA}

\begin{abstract}

We present high-quality, Keck spectroscopic data for a sample of 20
globular clusters (GCs) in the massive E0 galaxy NGC\,1407. A subset
of twenty line-strength indices of the Lick/IDS system have been
measured for both the GC system and the central integrated star-light
of the galaxy. Ages, metallicities and [$\alpha$/Fe] ratios have been
derived using several different approaches. The majority GCs in
NGC\,1407 studied are old, follow a tight metallicity sequence
reaching values slightly above solar, and exhibit mean [$\alpha$/Fe]
ratios of $\sim 0.3$\,dex. In addition, three GCs are formally derived
to be young ($\sim 4$\,Gyr), but we argue that they are actually old
GCs hosting blue horizontal branches.  We report, for the first time,
evidence for the existence of two chemically-distinct subpopulations
of metal-rich (MR) GCs. We find some MR GCs exhibit significantly
larger [Mg/Fe] {\it and} [C/Fe] ratios.  Different star formation
time-scales are proposed to explain the correlation between Mg and C
abundances. We also find striking CN overabundances over the entire GC
metallicity range. Interestingly, the behavior of C and N in
metal-poor (MP) GCs clearly deviates from the one in MR GCs. In
particular, for MR GCs, N increases dramatically while C essentially
saturates. This may be interpreted as a consequence of the increasing
importance of the CNO cycle with increasing metallicity.

\end{abstract}

\keywords{globular clusters: general --- galaxies: clusters ---
galaxies: formation}

\section{Introduction}

The formation of globular clusters (GCs) is thought to be linked to
major episodes of star formation in the galaxies (e.g.~Larson
1996; Elmegreen \& Efremov 1997; Ashman \& Zepf 2001). This fact is
supported by the systematic detection of young, massive star clusters
--presumably progenitors of present-day GCs-- in nearby, massive
star-forming regions in the Large Magellanic Cloud (van den Bergh
1994) and other galaxies like the Antennae (Whitmore \& Schweizer
1995), NGC\,1275 (Holtzman et al.~1992), NGC\,1569 (Ho \& Filippenko
1996) and NGC\,7252 (Miller et al.~1997). Therefore, observations of
extragalactic GCs are an essential tool to our understanding of how
the main star formation episodes of their host galaxies took place.

One of the most striking observational keys on the current topic is
the existence of bimodal color distributions in the GC systems of most
galaxies. This fact has been widely interpreted as an evidence for two
distinct GC subpopulations --metal-rich (MR; red) and metal-poor (MP;
blue), qualitatively resembling the two GC subpopulations of
spectroscopically-confirmed, different metallicities in our Galaxy--,
although in the recent work of Yoon, Yi \& Lee (2006) it is proposed
that such a bimodality could instead be driven by non-linearities in
the color-metallicity relationship. It is therefore clear that, if GC
metallicity subpopulations indeed exist --see e.g. Strader, Brodie \&
Beasley (2007) for NGC\,4472--, any convincing galaxy formation
scenario must be able to reproduce such a common phenomenon.

In this sense, three GC formation scenarios have been proposed so far
to explain the existence of GC subpopulations. In the {\sl major
merger} model (Ashman \& Zepf 1992; Zepf \& Ashman 1993), Es might be
the by-product of {\sl wet} (gas-rich) mergers of spiral galaxies
(Toomre \& Toomre 1972; Toomre 1977) with MR GCs being formed during
the merger process (Schweizer 1987), whereas the MP GC subpopulation
would be donated by the merging spirals. The {\sl in-situ} scenario
(Forbes et al.~1997) considers that the whole GC population was formed
within the host galaxy during a multi-phase dissipational
collapse. Bimodality arises as MP GCs would form at a very early epoch
whilst, after a dormant period of GC formation, MR GCs would form
slightly later along with the bulk of the galaxy stars. Feedback
processes (Forbes et al.~1997) and cosmic reionization (Santos 2003)
have been proposed as potential candidates to justify the truncation
of GC formation. Finally, based on the well-known relationship between
the mean metallicity of the GC system and the mass of the host galaxy
(Brodie \& Huchra 1991), the {\sl accretion} scenario (C\^ot\'e,
Marzke \& West 1998; C\^ot\'e et al.~2000; C\^ot\'e, West \& Marzke
2002) relies on the assumption that the MR GC subpopulation was formed
within a massive, seed, host galaxy whereas MP GCs were acquired as a
consequence of the accretion of their low-mass, host galaxies.

An essential prediction of the above GC formation scenarios deals with
the ages of both GC subpopulations. In all cases, the MP GC
subpopulation is predicted to be old. In the {\sl major merger} model,
the MR GC subpopulation is supposed to have the same age as that of
the main merger driving their formation. The {\sl in-situ/multiphase}
scenario predicts MR GCs to be just slightly younger (by $\sim 2 -
3$\,Gyr) than MP GCs, whereas no systematic age differences between GC
subpopulations are predicted by the accretion picture, as both MP and
MR GCs would originally form at high redshift in their respective host
galaxies before the accretion processes take place. Trying to
determine whether systematic differences in the ages of GC
subpopulations exist is, therefore, a crucial test to constrain the
reliability of the different GC formation scenarios.

As a consequence, during the last two decades there has been a large
effort to characterize the global properties and stellar populations
of extragalactic GCs. The arrival of 8 - 10\,m class telescopes has
made it possible to extend studies of GCs to a wide variety of galaxy
types, including massive ellipticals (Es), S0s, spiral galaxies and
dwarf Es (dEs). We refer the reader to Brodie \& Strader (2006) for a
thorough review of the present status of extragalactic GCs and their
implications for the current scenarios of galaxy formation and
assembly. Overall, most GCs in all types of galaxies seem to be {\it
old}, although it is worth noting that age-dating extragalactic GCs is
unavoidably subject to the poor age-disentangling power of single
stellar population (SSP) model predictions at the old regime.

In fact, all the above scenarios of GC formation may probably be valid
to some extent: there exist evidences for the existence of young GCs
in the {\it merger} remnant elliptical NGC\,3610 (Strader et
al.~2004), as well as hints for {\it accreted}, Galactic GCs
(hereafter GGCs) associated to the streams of Sagittarius (Layden \&
Sarajedini 2000; Bellazzini et al.~2003a,b) and Fornax-Leo-Sculptor
(Majewski 1994). Therefore, rather than questioning the reliability of
the three different scenarios, the debate should be focused on
determining the one that governs the formation of the bulk of GCs in
the Universe. In practice, answering this question might not be so
immediate either as the observational constrains of the different
scenarios would have blurred if they had happened at high redshift. In
this sense, on the basis of the semi-analytic galaxy formation model
by Cole et al.~(2000), Beasley et al.~(2002) adopted some elements of
the three classic scenarios to study GC bimodality in a hierarchical
galaxy formation context. More recently, supported by the old ages
inferred from direct comparison with GGCs for a high-quality,
spectroscopic subsample of GCs in eight Es, Strader et al.~(2005)
proposed a new GC formation picture that synthesizes aspects of both
accretion and in-situ approaches in the context of galaxy formation
through hierarchical merging at high-redshift (see also Rhode et
al.~2005). Nevertheless, the present topic is controversial as Puzia
et al.~(2005) reported that up to one third of the GCs in a sample of
seven early-type galaxies could have intermediate, spectroscopic ages
in the range $5 - 10$\,Gyr.

Actually, despite the observational efforts of the last years, the
current topic still lacks from a well-defined, high-quality,
representative sample of GCs in all types of galaxies. With GC systems
roughly following a {\it universal}, gaussian-like, luminosity
function that peaks at $M_{\rm V} \sim -7.5$ (e.g.~Harris 2001), the
use of 8 - 10\,m class telescopes for spectroscopic studies of GC
systems in galaxies beyond the Local Group is strongly demanded. Even
so, with the available instrumentation, only a subsample of a few
tenths --out of the brightest (say $B \leq 23$) GCs-- in a single
galaxy are subject to be reasonably observed for stellar population
diagnostics in, typically, 1--2 observing nights. As a consequence,
the problem of age-dating GC subpopulations is still far from being
addressed with the accuracy needed to derive conclusive results,
specially if we take into account that our current knowledge is still
biased to a few --the brightest-- GCs in a few --the closest--
galaxies. It is therefore clear that larger samples of GCs in much
larger samples of galaxies over a wide variety of types are necessary
to face, from a more robust, statistical point of view, the problem of
GC formation in the Universe.

The present paper on the GC system of NGC\,1407 is just intended to be
a step forward in the, presumably, long process of understanding the
stellar populations of extragalactic GCs, and their conclusions do not
necessarily have to be generalizated to other galaxies. With an
absolute magnitude in $K-$band of $M_{\rm K} \sim -25.6$ (Jarrett et
al.~2000), NGC\,1407 is a massive E0 dominating the Eridanus region of
galaxies (e.g.~Trentham et al.~2006; Brough et al.~2006) that may be
the progenitor of a brightest cluster galaxy (BCG). Its current
location at the centre of the group potential well, and intra-group
medium (O'Sullivan et al.~2001), may give rise to a different
evolutionary history than other group/cluster ellipticals. Indeed
recent simulations of BCGs (de Lucia et al.~2006) suggests an unique
hierarchical assembly history, with half of the final mass being being
added after redshifts of z $\sim 0.5$ (i.e. the last 5\,Gyrs). If this
recent assembly has been dissipational (i.e. involved gas) then we
might expect to find some signatures in the properties of the GC
system. 

In this paper we present unprecedented, high-quality spectroscopic
data for a subsample of GCs in NGC1407. Section~\ref{data} is devoted
to the GC sample selection, observations and data reduction. An
analysis of the kinematics of the GC subsample is presented in
Section~\ref{radvel}. Section~\ref{lick} describes the measurement of
Lick/IDS, line-strength indices (Gorgas et al~1993; Worthey et
al.~1994, hereafter W94; Worthey \& Ottaviani 1997, hereafter W97),
the ones are employed in Section~\ref{ages} to determine ages,
metallicities and abundance ratios of $\alpha$-elements for the galaxy
and the GC spectra using different approaches. A qualitative
description of the behavior of C and N sensitive Lick indices in our
GC sample is presented in Section~\ref{CN}. Section~\ref{discussion}
is reserved to discuss the most relevant results obtained in the
previous sections, and a general overview of the paper is finally
given in Section~\ref{summary}.

\section{Sample selection, observations and data reduction}
\label{data}

\subsection{GC sample selection}
\label{sample}

GC candidates of NGC\,1407 were selected on the basis of photometric
data taken in the F435W and F814LP filters using the Advanced Camera
for Surveys (ACS) in Wide Field Channel mode (WFC) on the {\sl Hubble
Space Telescope} (HST; proposal ID = 9427). We refer the reader to the
papers by Harris et al.~(2006) and Forbes et al.~(2006; hereafter F06)
for a thorough analysis of the above data.

\begin{figure*}[!t]
\center{
\scalebox{0.9}{\includegraphics{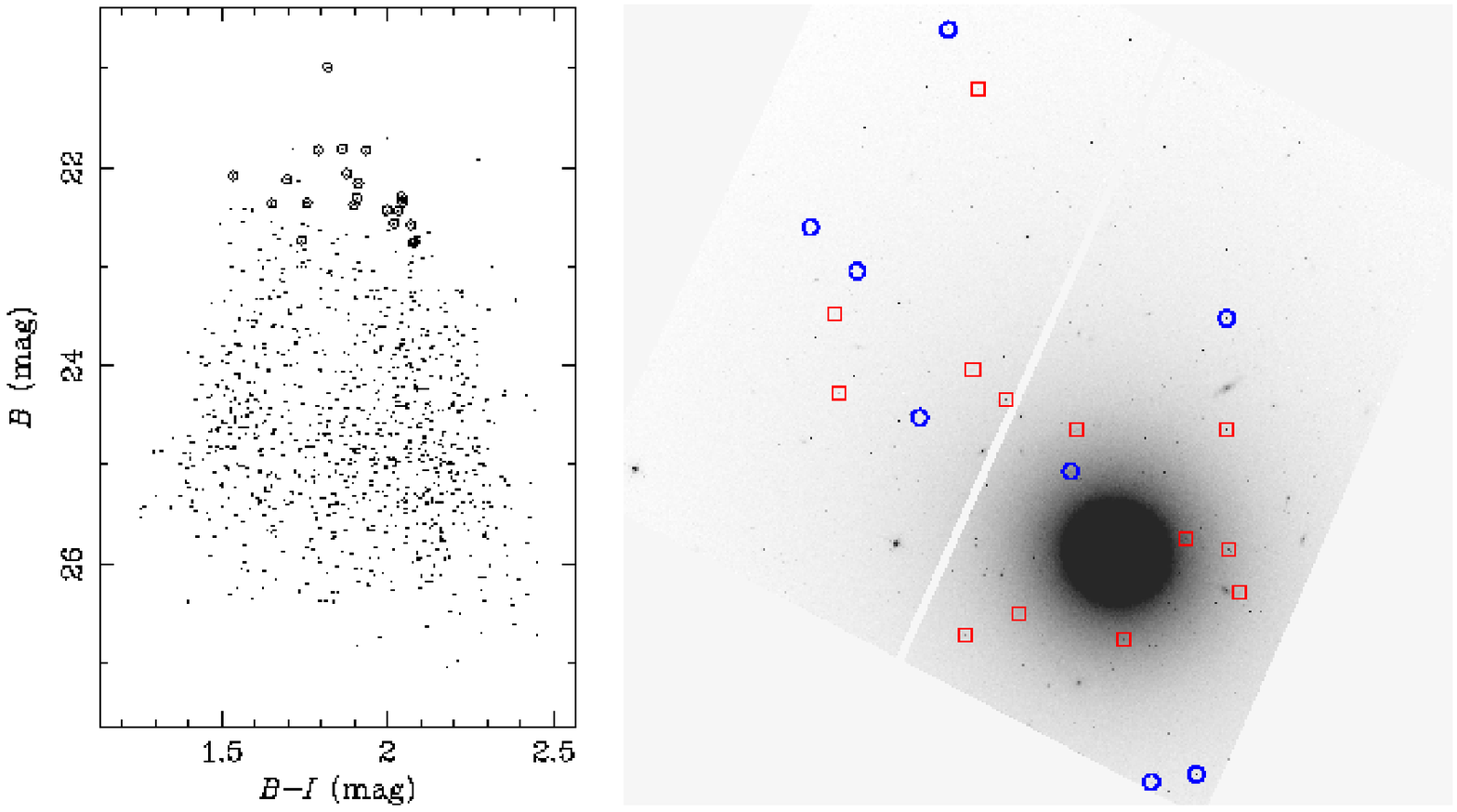}}
\caption{{\it Left:} color-magnitude diagram derived from HST ACS/WFC
photometry (Forbes et al.~2006) for the globular cluster candidates of
NGC\,1407. Each point in the diagram corresponds to a single GC
candidate. The 21 circled points indicate those GC candidates that
were selected for spectroscopy. {\it Right:} ACS/WFC imaging of
NGC\,1407 illustrating the location around the galaxy of the 21 GC
candidates for which spectroscopic data were obtained. Circles and
squares are displayed for $B - I < 1.87$ and $B - I > 1.87$
respectively.}
\label{fig1}
}
\end{figure*}

Figure~\ref{fig1} illustrates the $B$ versus $B-I$, color-magnitude
diagram (CMD) of the GCs candidates on the basis of the photometric
data given by F06 (left panel), as well as the corresponding ACS/WFC
imaging of NGC\,1407 in the F814LP filter (right panel). As expected
for a massive E galaxy, the GC system of NGC\,1407 exhibits a clear
bimodality in its color distribution, with red and blue subpopulations
peaking at $B-I \sim 1.61$ and $2.06$ respectively
(F06). Nevertheless, as discussed in Harris et al., the
high-luminosity regime ($B \la 23.5$) of this CMD does not exhibit an
obviously bimodal distribution. There exists a clear tilt in the
color-magnitude relation of the blue GC subpopulation that makes the
high-luminosity, blue GCs occupy the position of the bright end of the
red GC peak. Such a phenomenon, first reported by Ostrov et al.~(1998)
for the GC system of NGC\,1399 in Fornax (see also Dirsch et
al.~2003), has been detected for M87 and NGC\,4649 in Virgo (Strader
et al.~2006) and seems to be commonplace for bright Es dominating the
central regions of groups or clusters of galaxies (Harris et al.). A
mass-metallicity relationship, presumably though self-enrichment, has
been proposed in the above papers as a plausible explanation for this
phenomenon. Spectroscopic metallicities will help to test this
proposal.

From the high-luminosity regime of the CMD in Fig.~\ref{fig1}, 21 GC
candidates spanning a wide color range were selected for spectroscopic
purposes (circled points). It is worth noting that, as discussed by
Harris et al., some of the luminous objects we observed could be
remnant nuclei of disrupted, accreted dEs, although our subsequent
analysis finds no particular properties supporting the above
hypothesis. Because of the tilt of the blue sequence, the separation
between blue and red GC subpopulations is not well defined.
Throughout this paper, we have considered a tentative division at $B -
I = 1.87$ when it has been necessary.  Basic parameters of the
selected GC candidates, such as coordinates, colors, magnitudes and
heliocentric radial velocities are listed in Table~\ref{basicdata}.

\subsection{Observations and data reduction}
\label{obsdatared}

Spectroscopic data for 21 GC candidates in NGC\,1407 were obtained on
2003 October $21 - 23$ at the Keck~I telescope using the
Low-Resolution Imaging Spectrometer (LRIS; Oke et al.~1995) in
multi-object mode. With seeing varying from 0.6 to 0.8\,arcsec over
the three nights, all the exposures were made through a slitmask with
0.8\,arcsec width slits of $\sim 8$\,arcsec typical length. The total
exposure time for the GC candidates was 7 hours ($14\times30$\,min
exposures). In addition, a 15\,min ($3\times5$\,min) spectrum of
NGC\,1407 was obtained through a 0.8\,arcsec width long-slit
positioned along the major axis of the galaxy (P.A. = 56$^{o}$; Longo
et al.~1994, although note that it is not particularly well defined
for an E0 galaxy). In order to transform our data to the Lick/IDS
spectro-photometric system, a subsample of five K -- G giant stars
(HR\,2002, HR\,2153, HR\,2459, HR\,7429 and HR\,8165) from the
Lick/IDS stellar library (W94; W97) were also observed through the
long-slit mask during twilights. Short exposure times of $\sim 1$\,sec
were required to avoid saturation in their spectra. In turn, three
spectro-photometric standard stars from Oke (1990) were observed at
different air masses for calibrating purposes.

Making use of LRIS-B together with a dichroic splitting at 6800\,\AA,
spectra in the ranges $\sim3400-5800$\,\AA\ (blue) and
$\sim7100-9100$\,\AA\ (red) were simultaneously obtained for both the
GCs\footnote{Shifts in the GC spectral ranges of up to $\sim \pm
300$\,\AA\ with respect to the above mean values were obtained because
of the location of the different slitmasks along the dispersion
direction within the mask} and the galaxy. On the blue side, a
600\,line\,mm$^{-1}$ grism blazed at 4000\,\AA\ provided a reciprocal
dispersion of 0.6\,\AA~pix$^{-1}$ and an overall spectral resolution
(full width at half maximum; FWHM) of ~3.7\AA. The analysis of the red
side data is deferred to a forthcoming paper.
 
For each of the 14 GC mask frames, bi-dimensional (2D) distortions
were mapped by fitting polynomials to the edges of the individual
slitmasks. The raw mask frames, as well as their corresponding
overheads (flatfield and arc frames), were straightened by using the
derived polynomials, thus allowing the different slitmask regions in
each mask frame to be properly extracted into a single, rectangular
frame. After this, a standard spectroscopic reduction procedure (bias
subtraction, flatfielding, cosmic ray cleaning, C-distortion
correction, wavelength calibration, S-distortion, sky subtraction,
extinction correction, flux calibration, and final spectrum
extraction) was performed for each single slitmask with \reduceme
\footnote{http://www.ucm.es/info/Astrof/software/reduceme/reduceme.html}
(Cardiel~1999), as well as for the long-slit spectra of the galaxy and
the calibrating stars. The major improvement of this reduction package
is that it allows a parallel treatment of data and error frames along
the reduction procedure, thus producing an associated, reliable error
spectrum for each individual data spectrum.

In order to optimize the total exposure time of our targets,
wavelength calibrating arcs were only taken right after one of the
various GC mask exposures in each night. Because of flexures, small
shifts in both spatial and spectral directions were measured among
different GC mask frames. While spatial flexures did not remain a 
problem after flatfielding and extracting the slitmask regions into
single frames, the existence of spectral zero-point shifts among GC mask
frames could in principle prevent a proper wavelength calibration for all the
GC mask frames. In order to avoid this, for each slitmask of frames
without its own arc, spectral offsets with respect to the reference
slitmask were measured by cross-correlating their spectra around the
bright O\,{\sc i} skyline at $\sim 5577$\AA. After correcting the
slitmask zero-points, a standard wavelength
calibration for all the GC mask frames could finally be performed on
the basis of the available arcs.

The overall sky and galaxy background level in each GC mask frame was
computed by means of a standard, linear interpolation along the
spatial direction of the signal measured at regions free from the
contamination with the object of interest.  Special care was taken for
the shortest GC slitlets --of just $\sim 4 - 5$\,arcsec-- as only
a few pixels near the edges of the slits could be safely used for this
purpose. Moreover, GC mask frames \#06 and \#13 exhibited a rapidly
changing galaxy background since they were located in close proximity
to the galaxy. In these two cases, rather than linear fits, higher
order ($\leq 3$) polynomial fits to the background levels were
performed.

\begin{table*}[!t]
\centering{
\caption{Basic data for GC candidates in NGC\,1407}
\label{basicdata}
\begin{tabular}{cccccr}
\hline
Slit ID &  RA(J2000) & DEC(J2000) & $B$\,(mag)  & $B-I$\,(mag) & RV\,(km\,s$^{-1}$)\tablenotemark{a}\\
\hline
\#01 & 03:40:09.42  & $-$18:33:37.4  &$20.98\pm0.02$ & $1.82\pm0.03$ & $1196\pm 2(23)$\\
\#03 & 03:40:11.60  & $-$18:35:13.1  &$21.83\pm0.02$ & $1.93\pm0.03$ & $2204\pm 4(29)$\\
\#04 & 03:40:18.14  & $-$18:33:10.3  &$21.81\pm0.02$ & $1.86\pm0.03$ & $1717\pm 3(19)$\\
\#05 & 03:40:14.74  & $-$18:33:52.6  &$22.06\pm0.02$ & $1.88\pm0.03$ & $1669\pm 3(20)$\\
\#06 & 03:40:11.00  & $-$18:35:55.5  &$21.83\pm0.02$ & $1.79\pm0.03$ & $1379\pm 3(25)$\\
\#07 & 03:40:17.56  & $-$18:33:59.6  &$22.16\pm0.02$ & $1.91\pm0.03$ & $2055\pm 4(18)$\\
\#08 & 03:40:12.69  & $-$18:34:23.0  &$22.12\pm0.02$ & $1.70\pm0.03$ & $1843\pm 5(23)$\\
\#09 & 03:40:17.16  & $-$18:33:23.4  &$22.08\pm0.02$ & $1.53\pm0.03$ & $1642\pm 8(26)$\\
\#11 & 03:40:09.38  & $-$18:34:46.0  &$22.31\pm0.02$ & $1.91\pm0.03$ & $1652\pm 6(30)$\\
\#12 & 03:40:09.16  & $-$18:34:59.2  &$22.35\pm0.02$ & $2.05\pm0.03$ & $2254\pm 5(22)$\\
\#13 & 03:40:10.06  & $-$18:35:53.2  &$22.36\pm0.02$ & $1.76\pm0.03$ & $1558\pm 6(26)$\\
\#14 & 03:40:14.90  & $-$18:35:11.9  &$22.30\pm0.02$ & $2.04\pm0.03$ & $1947\pm 4(21)$\\
\#15 & 03:40:09.43  & $-$18:34:10.4  &$22.38\pm0.02$ & $1.90\pm0.03$ & $1636\pm 4(19)$\\
\#17 & 03:40:14.61  & $-$18:32:29.2  &$22.43\pm0.02$ & $2.03\pm0.03$ & $1920\pm 3(22)$\\
\#18 & 03:40:12.55  & $-$18:34:10.6  &$22.43\pm0.02$ & $2.00\pm0.03$ & $2402\pm11(23)$\\
\#22 & 03:40:14.04  & $-$18:34:01.6  &$22.58\pm0.02$ & $2.07\pm0.04$ & $1547\pm 3(17)$\\
\#26 & 03:40:15.25  & $-$18:32:11.5  &$22.36\pm0.02$ & $1.65\pm0.03$ & $1907\pm 6(30)$\\
\#27 & 03:40:17.64  & $-$18:33:35.9  &$22.75\pm0.02$ & $2.08\pm0.03$ & $1839\pm 5(26)$\\
\#30 & 03:40:10.28  & $-$18:34:42.8  &$22.56\pm0.02$ & $2.02\pm0.04$ & $1287\pm 7(34)$\\
\#31 & 03:40:13.76  & $-$18:35:05.3  &$22.76\pm0.02$ & $2.08\pm0.03$ & $  62\pm 4(21)$\\
\#33 & 03:40:15.85  & $-$18:34:06.9  &$22.74\pm0.02$ & $1.74\pm0.03$ & $1732\pm 9(22)$\\
\hline
\end{tabular}
} \tablenotetext{a}{Final heliocentric radial velocities of the GC
candidates together with their random and absolute errors (in
parentheses; see text).  Candidate \#31 was found to be a foreground
Galactic star.}
\end{table*}

\section{Radial velocity measurements and GC kinematics}
\label{radvel}

Radial velocities for the GC candidates were derived using the MOVEL
algorithm described in Gonz\'alez~(1993), which is an improvement of
the classic Fourier quotient method by Sargent et al.~(1977) to
determine RVs and velocity dispersions of galaxy spectra. Although the
spectral resolution of our data does not allow us to measure reliable
velocity dispersions for the GCs, we preferred to use this method given
that, from previous experience, we know it provides a more robust RV
determination than classical cross-correlation procedures. In order to
prevent uncertainties in the method that appear when velocity
dispersion solutions are very close to zero, the GC spectra were
broadened by convolving them with a Gaussian function of $\sigma =
50$\,km\,s$^{-1}$.

The five Lick/IDS stars observed during the run were separately used
as reference spectra for this method, thus deriving five independent
values of RVs for each GC. In each case, random errors in the derived
RVs were computed by MonteCarlo simulations in which Gaussian noise is
added to the object and reference spectra according to their
corresponding error spectra. Final values for radial velocities were
computed as error-weighted means of the five individual estimations,
with the larger of either the typical error of the error-weighted mean
or the error-weighted standard deviation of the single determinations
being considered as the final random error. It is important to note
that such final random errors do not account for any systematic effect
arising, for example, from uncertainties in the wavelength
calibration. In other words, the derived RVs and their errors are
fully consistent within their own system, but small offsets could
still be affecting the true values of the radial velocities. In order
to account for errors due to small deviations from a linear wavelength
scale, the GC spectra were also cross-correlated with an appropriate
{\sl MILES} \footnote{Medium resolution INT Library of Empirical
Spectra; http://www.ucm.es/info/Astrof/miles/miles.html}
(S\'anchez-Bl\'azquez et al.~2006a; Cenarro et al.~2007) template in
$\sim 150$ small, overlapping spectral regions along wavelength. The
{\sl rms} standard deviation of all the RV solutions obtained from
each individual region can be considered as a more reliable estimation
of the absolute uncertainties of the RVs. The final heliocentric
radial velocities of the GC candidates, their corresponding random
errors and their absolute errors (within parenthesis) are listed in
last column of Table~\ref{basicdata}. Candidate \#31 turned out to be
a foreground Galactic star that was rejected from further analysis in
this paper.

The above procedure was followed to measure the kinematics of the
integrated central spectrum of NGC\,1407 (R$_{\rm eff}$/8 aperture).
An R$_{\rm eff}$ of 36.4\,arcsec for NGC\,1407 was taken from Diehl \&
Statler (2006). The derived central velocity dispersion and the
heliocentric radial velocity are $293.5 \pm 4.1$\,km\,s$^{-1}$ and
$1785.6 \pm 5.2$\,km\,s$^{-1}$ respectively.  These values are in
reasonably good agreement with previous determinations as given in
Hyperleda, i.e. 273\,km\,s$^{-1}$ and $1783 \pm 26$\,km\,s$^{-1}$
respectively.

On the basis of the 20 bona fide GCs, the mean heliocentric radial
velocity of our GC subsystem (RV$_{\rm sys}$) is $1769 \pm
70$\,km\,s$^{-1}$, with an overall {\sl rms} standard deviation of
313\,km\,s$^{-1}$. Figure~\ref{RV-r} displays the distribution of such
GC RVs as a function of the projected distance from the center of
NGC\,1407 (galactocentric radius; $R_{\rm gc}$). A distance of
20.9\,Mpc to NGC\,1407 (F06) has been assumed for this figure (note
that the galaxy could be as far as 28.8\,Mpc; Tonry et al.~2001). In
any case, it is clear that the dispersion of the GC RVs seems to
decrease with the increasing projected distance. In particular, when
the {\sl rms} standard deviation of the GC RVs is separately computed
within three different radial bins, we obtain the following values:
361\,km\,s$^{-1}$ ($R_{\rm gc} < 10$\,kpc), 244\,km\,s$^{-1}$ ($5 <
R_{\rm gc} < 15$\,kpc) and 149\,km\,s$^{-1}$ ($R_{\rm gc} > 10$\,kpc).
This behaviour could have important implications for orbital
properties of the GCs within the potential well of the
galaxy. Although our GC sample is not large enough to make definitive
statements on this issue, the existence of radial orbits (rather than
isotropic ones) would help to explain the distribution of GC RVs shown
in Fig.~\ref{RV-r}. A more detailed analysis on the GC orbits, galaxy
mass and dark matter content estimates are deferred to a forthcoming
paper.

Rotation around the host galaxy is a common signature of GC systems,
although there exists no well-established pattern for massive
Es. Whilst some GC systems do not exhibit any rotation (NGC\,1399;
Richtler et al.~2004), others show significant rotation (M87, C\^ot\'e
et al.~2001; NGC\,4472, Zepf et al.~2000; C\^ot\'e et al.~2003). In
Figure~\ref{checkrot} we illustrate the GC radial velocities relative
to RV$_{\rm sys}$ as a function of the projected distance to four
different directions: the major and minor axes, and the two
intermediate orientations. It is clear from the figure that there
exist no obvious rotation around any direction. There may still be a
weak rotating signature for MP GCs although, once again, the small
number of GCs does not allow us to carry out a reliable statistical
analysis.

\section{Measurement of Lick indices}
\label{lick}

The Lick/IDS system of line-strength indices (Gorgas et al.~1993; W94)
has been widely used as the reference system to interpret the stellar
population properties in the optical spectral range of a large variety
of stellar systems (see e.g.~Trager et al.~2000a,b;
S\'anchez-Bl\'azquez et al.~2006b and references therein). It is worth
mentioning though that, during the past few years, there have been
significant efforts in developing much improved empirical stellar
libraries (e.g.~the Indo-US stellar library, Vald\'es et al.~2004;
{\sl MILES}, S\'anchez-Bl\'azquez et al.~2006a, Cenarro et al.~2007)
that comprise larger samples of stars, over wider ranges of
atmospheric parameters and with a higher spectral resolution than the
Lick stellar sample. In the near future, these works are expected to
constitute the new reference system for forthcoming studies on this
topic. SSP model predictions on the basis of such libraries
(e.g.~Vazdekis et al.~2007, in preparation) are expected to overcome
some limitations of the current ones based on the Lick stellar
library.

\begin{deluxetable}{lccc@{}cr@{}r}
\tablecolumns{7}
\tablewidth{0pt} \tablecaption{List of Lick/IDS indices measured in this work.\label{Lickset}}
\tablehead{
\colhead{$I$\tablenotemark{a}} & \colhead{$\sigma$\,(km\,s$^{-1}$)\tablenotemark{b}} & \colhead{$\Delta I$\tablenotemark{c}} &           \multicolumn{2}{c}{Errors}                                                          &    \multicolumn{2}{c}{$t$-tests} \\
\colhead{          } & \colhead{                                         } & \colhead{                       } & \colhead{$\sigma_{\rm typ}$\tablenotemark{d}} &\colhead{$\sigma_{\rm rms}$\tablenotemark{d}} & \colhead{$t_{\rm typ}$\tablenotemark{f}} & \colhead{$t_{\rm rms}$\tablenotemark{g}} \\
}
\startdata
 H$\delta_{A}$(\AA)   &325 & $-0.075$ &  0.070\ \ \ &  0.097 &  1.1\ \ \ &  0.8\\
 H$\delta_{F}$(\AA)   &325 & $-0.020$ &  0.058\ \ \ &  0.052 &  0.4\ \ \ &  0.4\\
 CN$_{1}$(mag)        &325 & $-0.011$ &  0.002\ \ \ &  0.002 &  5.3\ \ \ &  5.2\\
 CN$_{2}$(mag)        &325 & $-0.016$ &  0.002\ \ \ &  0.005 &  6.8\ \ \ &  3.6\\
 Ca4227(\AA)          &300 & $+0.034$ &  0.028\ \ \ &  0.035 &  1.2\ \ \ &  1.0\\
 G4300(\AA)           &300 & $-0.053$ &  0.052\ \ \ &  0.112 &  1.0\ \ \ &  0.5\\
 H$\gamma_{A}$(\AA)   &275 & $-0.528$ &  0.042\ \ \ &  0.031 & 12.7\ \ \ & 17.2\\
 H$\gamma_{F}$(\AA)   &275 & $-0.233$ &  0.033\ \ \ &  0.033 &  7.0\ \ \ &  7.0\\
 Fe4383(\AA)          &250 & $-0.181$ &  0.076\ \ \ &  0.031 &  2.4\ \ \ &  5.9\\
 Ca4455(\AA)          &250 & $-0.302$ &  0.072\ \ \ &  0.059 &  4.2\ \ \ &  5.1\\
 Fe4531(\AA)          &250 & $-0.377$ &  0.073\ \ \ &  0.054 &  5.2\ \ \ &  6.9\\
 Fe4668(\AA)          &250 & $+0.357$ &  0.077\ \ \ &  0.040 &  4.6\ \ \ &  9.0\\
 H$\beta$(\AA)        &225 & $+0.053$ &  0.025\ \ \ &  0.012 &  2.1\ \ \ &  4.3\\
 Fe5015(\AA)          &200 & $-0.022$ &  0.052\ \ \ &  0.061 &  0.4\ \ \ &  0.4\\
 Mg$_{1}$(mag)        &200 & $-0.004$ &  0.001\ \ \ &  0.002 &  5.9\ \ \ &  1.6\\
 Mg$_{2}$(mag)        &200 & $-0.010$ &  0.001\ \ \ &  0.002 & 11.5\ \ \ &  4.7\\
 Mg$b$(\AA)           &200 & $+0.074$ &  0.027\ \ \ &  0.022 &  2.8\ \ \ &  3.4\\
 Fe5270(\AA)          &200 & $+0.210$ &  0.030\ \ \ &  0.018 &  7.0\ \ \ & 11.5\\
 Fe5335(\AA)          &200 & $+0.223$ &  0.029\ \ \ &  0.039 &  7.8\ \ \ &  5.7\\
 Fe5406(\AA)          &200 & $+0.087$ &  0.030\ \ \ &  0.041 &  2.9\ \ \ &  2.1\\
\enddata
\tablecomments{(a) Lick index; (b) Overall spectral resolution at which the index was
measured; (c) Error-weighted, index offset between the
spectro-photometric system of this work ($TW$) and the corresponding
Lick system (W94; W97): $\Delta I = I(TW) - I(W94/W97)$; (d) Typical error of the offset
$\Delta I$; (e) Error-weighted, {\sl rms}, residual standard deviation
of the offset $\Delta I$; (f, g) Values of $t$-tests using
$\sigma_{\rm typ}$ and $\sigma_{\rm rms}$ as estimates of the offset
uncertainties respectively. Assuming a significance level of $\alpha =
0.1$ and 4 ($5 - 1$) degrees of freedom, the critical value for the
bilateral test is $t_{\alpha/2,}$$_{\,4} = 2.132$.}
\end{deluxetable}

We have measured all the Lick indices within the spectral
range of the GCs and the galaxy spectra. In order to compare our index
measurements with those predicted by SSPs models in the Lick system,
we matched both the spectral resolution and the
spectro-photometric system of the models. Since the spectral resolution
of the Lick stellar library varies with wavelength, each single index
had to be measured at a given spectral resolution that, in all cases,
was much lower than the overall spectral resolution of the GC spectra
($\sim 100$\,km\,s$^{-1}$).  Hence, before measuring the indices, all
the GC spectra were broadened by convolving with an appropriate
Gaussian function to match the required spectral resolution. For the
galaxy spectrum, with an overall spectral resolution $\sim
300$\,km\,s$^{-1}$, we applied the sigma-dependent, polynomial
corrections given in Gorgas et al.~(2007; submitted).

As mentioned in Section~\ref{obsdatared}, the 5 stars in common
with the Lick stellar library were employed to correct our data to the
Lick spectro-photometric system. After broadening their spectra to
match the required spectral resolutions, the Lick indices of the stars
were measured and compared with those published in W94 and W97, as it
is illustrated in Figure~\ref{lickoffs}. For each index, an
error-weighted, mean offset was derived with respect to the Lick
system. The index errors estimated in W94 and W97 for a typical,
single-observation, star were assumed for HR\,2153, HR\,2459, and
HR\,8165 (filled circles). For the {\it standards} HR\,2002 and
HR\,7429 (open squares), with 41 and 64 repeated observations
respectively, their corresponding typical errors were instead
considered.

Even though all the derived offsets were systematically applied to our
data, we computed $t$-tests to check whether the derived calibrations
were statistically significant or not. For a significance level of
$\alpha = 0.1$ and 4 degrees of freedom (5 data points minus 1 for the
mean), the derived offsets can be considered statistically significant
if their correponding $t$-tests are larger than $\sim 2.1$. Note that
although the above calibrations rely on an error-weighted {\it
effective} number of stars in the range $1.6 - 3.0$ (depending on the
index), the derived offsets seem to be reasonably well constrained. In
fact, the two Lick {\it standard} stars --that essentially dominate
the calibrations because of their considerably smaller errors--
exhibit an excellent agreement in nearly all cases. The only possible
exception is G4300 which appears slightly less well constrained.

The Lick indices we have measured are listed in Table~\ref{Lickset},
together with the spectral resolutions at which the indices were
measured, the derived offsets w.r.t the Lick system, their errors and
their corresponding $t$-tests. To prevent possible underestimates of
the offset uncertainties arising from the small number of points, we
computed not only the typical error of the offsets ($\sigma_{\rm
typ}$) but also the error-weighted, {\it rms}, residual standard
deviation of the offsets ($\sigma_{\rm rms}$). Consequently, the
$t$-test values are provided for both error estimates ($t_{\rm typ}$
and $t_{\rm rms}$).

In addition to the above Lick indices, we have also computed the indices
\begin{equation}
{\rm [MgFe]'} = \sqrt{{\rm Mg}b\times(0.72 \times {\rm Fe5270}+0.28 \times {\rm Fe5335})},
\end{equation}
as defined by Thomas et al.~(2003a; hereafter TMB03), and
\begin{equation}
<{\rm Fe}> = ({\rm Fe5270}+{\rm Fe5335})/2.
\end{equation}
In Table~\ref{Lickindices} we present the final index measurements
for the GCs and the central galaxy spectrum.

\clearpage

\thispagestyle{empty}

\begin{deluxetable}{@{}l@{}c@{}r@{}r@{}r@{}r@{}r@{}r@{}r@{}r@{}r@{}r@{}r@{}r@{}r@{}r@{}r@{}r@{}r@{}r@{}r@{}r@{}r@{}r@{}}
\tablecolumns{24}
\tabletypesize{\tiny}
\rotate
\tablewidth{0pt}
\tablecaption{Lick indices of the GCs and the $R_{\rm eff}/8$ central
spectrum of NGC\,1407.\label{Lickindices}
}
\tablehead{
\colhead{Slit ID}&\colhead{$S/N$}&\colhead{H$\delta_{A}$}&\colhead{H$\delta_{F}$}&\colhead{CN$_{1}$}&\colhead{CN$_{2}$}&\colhead{Ca42}&\colhead{G43}&\colhead{H$\gamma_{A}$}&\colhead{H$\gamma_{F}$}&\colhead{Fe43}&\colhead{Ca44}&\colhead{Fe45}&\colhead{Fe46}&\colhead{H$\beta$}&\colhead{Fe50}&\colhead{Mg$_{1}$}&\colhead{Mg$_{2}$}&\colhead{Mg$b$}&\colhead{Fe52}&\colhead{Fe53}&\colhead{Fe54}&\colhead{[MgFe]'}&\colhead{$<$Fe$>$}\\*
\colhead{        }&\colhead{(\AA$^{-1}$)}&\colhead{(\AA)}&\colhead{(\AA)}&\colhead{(mag)}&\colhead{(mag)}&\colhead{(\AA)}&\colhead{(\AA)}&\colhead{(\AA)}&\colhead{(\AA)}&\colhead{(\AA)}&\colhead{(\AA)}&\colhead{(\AA)}&\colhead{(\AA)}&\colhead{(\AA)}&\colhead{(\AA)}&\colhead{(mag)}&\colhead{(mag)}&\colhead{(\AA)}&\colhead{(\AA)}&\colhead{(\AA)}&\colhead{(\AA)}&\colhead{(\AA)}&\colhead{(\AA)}
}
\startdata
\hline
\#01 &$56 - 93$  &     1.06&     1.60&     0.007&     0.047&     0.48&     3.94&  $-$2.33&     0.54&     3.05&     1.21&     2.41&     1.00&     2.08&     3.74&     0.050&     0.142&     2.28&     1.47&     1.24&     0.85&     1.79&     1.35\\
     &           &     0.13&     0.09&     0.004&     0.004&     0.07&     0.12&     0.13&     0.08&     0.17&     0.09&     0.13&     0.19&     0.08&     0.17&     0.002&     0.002&     0.09&     0.10&     0.11&     0.08&     0.06&     0.07\\
\#03 &$20 - 35$  &  $-$1.28&     0.59&     0.131&     0.175&     0.68&     5.10&  $-$4.11&  $-$0.57&     3.91&     1.59&     3.78&     2.07&     1.42&     4.30&     0.075&     0.208&     3.52&     1.57&     1.67&     1.40&     2.37&     1.62\\
     &           &     0.39&     0.27&     0.011&     0.013&     0.19&     0.31&     0.36&     0.22&     0.44&     0.23&     0.34&     0.51&     0.21&     0.44&     0.005&     0.006&     0.22&     0.25&     0.28&     0.21&     0.17&     0.19\\
\#04 &$31 - 53$  &  $-$0.01&     0.92&     0.022&     0.054&     0.54&     4.35&  $-$2.83&     0.11&     2.81&     1.43&     2.61&     1.89&     1.92&     3.77&     0.051&     0.155&     2.48&     1.53&     1.42&     0.83&     1.93&     1.47\\
        &        &     0.25&     0.17&     0.007&     0.008&     0.12&     0.21&     0.23&     0.14&     0.30&     0.16&     0.23&     0.34&     0.14&     0.29&     0.003&     0.004&     0.15&     0.17&     0.19&     0.14&     0.10&     0.13\\
\#05 &$22 - 41$  &     0.75&     1.50&     0.068&     0.124&     0.32&     4.11&  $-$2.74&     0.18&     3.64&     1.62&     3.77&  $-$0.37&     1.84&     5.15&     0.070&     0.205&     3.05&     2.07&     1.58&     1.47&     2.43&     1.83\\
        &        &     0.35&     0.24&     0.010&     0.011&     0.17&     0.28&     0.30&     0.18&     0.39&     0.21&     0.30&     0.46&     0.18&     0.38&     0.004&     0.005&     0.20&     0.22&     0.24&     0.18&     0.13&     0.16\\
\#06 &$28 - 49$  &     1.35&     1.74&     0.021&     0.069&     0.64&     3.78&  $-$1.45&     0.74&     2.37&     0.82&     3.69&     1.45&     1.93&     3.67&     0.035&     0.126&     2.00&     1.69&     1.35&     0.74&     1.79&     1.52\\
        &        &     0.26&     0.18&     0.007&     0.009&     0.13&     0.22&     0.24&     0.15&     0.33&     0.17&     0.24&     0.37&     0.15&     0.31&     0.003&     0.004&     0.16&     0.18&     0.21&     0.16&     0.11&     0.14\\
\#07 &$21 - 37$  &  $-$0.02&     1.26&     0.076&     0.101&     0.74&     4.23&  $-$2.30&     0.80&     3.47&     1.28&     2.05&     0.25&     1.89&     4.69&     0.066&     0.176&     2.93&     2.20&     1.85&     1.43&     2.48&     2.03\\
        &        &     0.37&     0.25&     0.010&     0.012&     0.18&     0.30&     0.33&     0.20&     0.43&     0.23&     0.33&     0.50&     0.19&     0.41&     0.005&     0.006&     0.22&     0.23&     0.26&     0.20&     0.14&     0.18\\
\#08 &$18 - 25$  &     1.53&     1.56&     0.012&     0.028&     0.64&     3.54&  $-$1.03&     1.25&     2.20&     0.57&     2.19&     0.90&     2.29&     4.27&     0.033&     0.107&     1.85&     0.93&     1.14&     1.13&     1.35&     1.03\\
        &        &     0.43&     0.30&     0.012&     0.015&     0.22&     0.38&     0.42&     0.26&     0.59&     0.31&     0.47&     0.69&     0.28&     0.60&     0.007&     0.008&     0.30&     0.36&     0.41&     0.30&     0.22&     0.27\\
\#09 &$26 - 37$  &     2.86&     2.42&  $-$0.048&  $-$0.012&     0.22&     2.63&     0.99&     2.03&     0.53&     0.80&     2.00&  $-$1.06&     2.43&     1.73&     0.015&     0.068&     0.48&     1.12&     0.96&     0.22&     0.72&     1.04\\
        &        &     0.28&     0.19&     0.008&     0.010&     0.15&     0.27&     0.27&     0.17&     0.41&     0.20&     0.31&     0.48&     0.19&     0.42&     0.004&     0.005&     0.22&     0.23&     0.27&     0.21&     0.17&     0.18\\
\#11 &$14 - 25$  &  $-$1.41&     0.43&     0.099&     0.136&     0.91&     5.14&  $-$3.71&  $-$0.58&     3.65&     1.97&     3.70&     1.57&     1.51&     5.25&     0.060&     0.177&     2.78&     2.02&     1.52&     1.25&     2.29&     1.77\\
        &        &     0.56&     0.38&     0.015&     0.018&     0.26&     0.44&     0.50&     0.31&     0.63&     0.32&     0.48&     0.74&     0.29&     0.62&     0.007&     0.008&     0.32&     0.35&     0.41&     0.30&     0.22&     0.27\\
\#12 &$13 - 25$  &  $-$2.11&     0.27&     0.139&     0.182&     0.39&     4.79&  $-$5.25&  $-$1.09&     4.55&     1.20&     3.58&     1.92&     1.63&     4.53&     0.091&     0.258&     4.17&     2.65&     2.45&     1.22&     3.29&     2.55\\
        &        &     0.61&     0.40&     0.016&     0.019&     0.28&     0.46&     0.52&     0.31&     0.62&     0.34&     0.47&     0.72&     0.29&     0.62&     0.007&     0.009&     0.32&     0.35&     0.38&     0.30&     0.21&     0.26\\
\#13 &$17 - 31$  &     1.57&     1.14&  $-$0.006&     0.041&     0.30&     3.87&  $-$1.95&     0.96&     3.37&     1.34&     2.32&     1.26&     2.02&     1.72&     0.035&     0.144&     2.30&     1.44&     1.51&     1.02&     1.83&     1.48\\
        &        &     0.43&     0.31&     0.012&     0.014&     0.21&     0.36&     0.39&     0.23&     0.50&     0.26&     0.40&     0.59&     0.23&     0.51&     0.005&     0.007&     0.26&     0.28&     0.32&     0.24&     0.17&     0.22\\
\#14 &$17 - 33$  &  $-$1.48&     0.68&     0.116&     0.162&     1.06&     5.80&  $-$5.81&  $-$0.51&     5.03&     1.66&     3.37&     2.98&     1.43&     3.12&     0.081&     0.228&     4.30&     2.01&     2.13&     1.31&     2.96&     2.07\\
        &        &     0.48&     0.32&     0.013&     0.015&     0.21&     0.35&     0.41&     0.23&     0.47&     0.26&     0.38&     0.55&     0.22&     0.49&     0.005&     0.007&     0.24&     0.27&     0.29&     0.22&     0.17&     0.20\\
\#15 &$16 - 30$  &  $-$1.05&     0.84&     0.075&     0.105&     0.78&     5.50&  $-$3.55&     0.19&     3.46&     1.71&     3.51&     4.05&     1.68&     4.28&     0.074&     0.197&     3.54&     1.53&     1.83&     0.66&     2.39&     1.68\\
        &        &     0.48&     0.32&     0.013&     0.015&     0.22&     0.36&     0.41&     0.25&     0.52&     0.27&     0.40&     0.58&     0.24&     0.52&     0.006&     0.007&     0.26&     0.29&     0.33&     0.26&     0.19&     0.22\\
\#17 &$18 - 37$  &  $-$1.45&     0.56&     0.136&     0.194&     0.80&     5.27&  $-$4.93&  $-$0.98&     4.05&     1.38&     3.35&     3.10&     2.19&     5.96&     0.102&     0.257&     4.12&     2.11&     1.87&     1.28&     2.90&     1.99\\
        &        &     0.44&     0.29&     0.012&     0.014&     0.20&     0.33&     0.38&     0.23&     0.46&     0.24&     0.34&     0.49&     0.20&     0.42&     0.005&     0.006&     0.22&     0.24&     0.27&     0.20&     0.16&     0.18\\
\#18 &$14 - 26$  &  $-$0.99&     0.09&     0.076&     0.105&     0.79&     4.64&  $-$4.08&  $-$0.89&     3.12&     1.81&     3.75&     1.41&     2.06&     3.85&     0.076&     0.214&     3.25&     2.00&     2.21&     1.47&     2.59&     2.11\\
        &        &     0.56&     0.38&     0.015&     0.018&     0.26&     0.44&     0.50&     0.31&     0.65&     0.33&     0.47&     0.73&     0.29&     0.63&     0.007&     0.008&     0.32&     0.35&     0.39&     0.29&     0.21&     0.26\\
\#22 &$12 - 26$  &  $-$2.88&  $-$0.29&     0.191&     0.229&     1.24&     4.53&  $-$4.98&  $-$1.00&     4.66&     1.66&     3.77&     2.91&     1.42&     3.58&     0.121&     0.273&     4.44&     2.53&     2.73&     1.69&     3.39&     2.63\\
        &        &     0.68&     0.45&     0.019&     0.022&     0.29&     0.49&     0.54&     0.32&     0.64&     0.35&     0.48&     0.70&     0.29&     0.62&     0.007&     0.009&     0.31&     0.33&     0.37&     0.28&     0.21&     0.25\\
\#26 &$22 - 34$  &     2.41&     2.14&  $-$0.034&  $-$0.003&     0.59&     3.43&  $-$1.08&     1.02&     2.59&     1.01&     2.33&     0.00&     2.26&     2.88&     0.039&     0.116&     1.57&     0.96&     0.78&     0.62&     1.19&     0.87\\
        &        &     0.34&     0.24&     0.010&     0.012&     0.18&     0.31&     0.33&     0.20&     0.45&     0.23&     0.35&     0.53&     0.21&     0.46&     0.005&     0.006&     0.24&     0.27&     0.31&     0.23&     0.17&     0.20\\
\#27 &$11 - 25$  &  $-$2.65&     0.90&     0.160&     0.209&     1.07&     5.33&  $-$5.85&  $-$1.18&     4.76&     1.49&     2.64&     5.44&     1.39&     4.96&     0.102&     0.281&     4.83&     2.49&     1.65&     1.78&     3.30&     2.07\\
        &        &     0.73&     0.46&     0.020&     0.023&     0.31&     0.52&     0.59&     0.34&     0.68&     0.37&     0.54&     0.72&     0.31&     0.65&     0.007&     0.009&     0.33&     0.36&     0.40&     0.30&     0.23&     0.27\\
\#30 &$ 8 - 14$  &  $-$0.72&     0.92&     0.065&     0.120&     1.20&     3.60&  $-$2.99&  $-$0.52&     3.46&     1.15&     4.02&     2.37&     1.34&     3.33&     0.049&     0.227&     4.12&     2.08&     1.73&     1.67&     2.86&     1.90\\
        &        &     1.02&     0.70&     0.028&     0.033&     0.45&     0.83&     0.88&     0.55&     1.12&     0.61&     0.81&     1.25&     0.51&     1.13&     0.012&     0.015&     0.52&     0.60&     0.69&     0.52&     0.39&     0.46\\
\#33 &$11 - 19$  &     3.90&     2.38&  $-$0.071&  $-$0.023&     1.34&     1.69&     0.56&     1.75&     2.53&     1.44&     3.07&     2.61&     2.57&     2.90&     0.052&     0.121&     2.09&     1.39&     1.16&     1.05&     1.67&     1.28\\
        &        &     0.65&     0.47&     0.019&     0.023&     0.33&     0.62&     0.61&     0.38&     0.84&     0.44&     0.65&     0.95&     0.38&     0.85&     0.009&     0.011&     0.43&     0.48&     0.54&     0.40&     0.29&     0.36\\
$R_{\rm eff}/8$ &$164 - 281$&  $-$3.17&  $-$0.17&     0.161&     0.207&     1.28&     5.45&  $-$6.46&  $-$1.92&     5.45&     2.14&     3.98&     8.17&     1.52&     6.18&     0.180&     0.355&     5.31&     3.01&     2.71&     1.85&     3.94&     2.86\\
                &           &     0.06&     0.10&     0.001&     0.004&     0.04&     0.09&     0.05&     0.04&     0.13&     0.20&     0.09&     0.09&     0.05&     0.20&     0.001&     0.001&     0.05&     0.04&     0.11&     0.07&     0.03&     0.06\\
\enddata
\tablecomments{Index errors accounting for photon statistic and uncertainties
in RVs are given below the index measurements. The indices $<$Fe$>$
and [MgFe]' are also included in the table.}
\tablenotetext{a}{Mean signal-to-noise ratios per angstrom of
the final spectra at the location of the bluest and reddest indices,
H$\delta_{A}$ and Fe5406.}
\end{deluxetable}

\clearpage

\section{Ages, metallicities and $\alpha$-element abundance ratios for the GCs}
\label{ages}

In this section we study the stellar population properties of the
GCs. For this purpose, we make use of the Lick index measurements
presented in the previous section along with independent approaches to
derive ages, metallicities and $\alpha$-element abundance ratios. In
Section~\ref{PCA}, a first estimate of the GC metallicities is
obtained on the basis of the metallicity calibration by Strader \&
Brodie~(2004). The other two approaches (Section~\ref{index-index})
rely on the SSP models of TMB03 and Thomas et al.~(2004; hereafter
TMK04), which account for three different [$\alpha$/Fe] ratios as well
as for blue horizontal branch (BHB) effects for metallicities $\leq
-0.5$\,dex (Maraston et al.~2003). Ages, metallicities and
[$\alpha$/Fe] ratios derived for the GCs and the central spectrum of
NGC\,1407 according to these different procedures are presented in
Table~\ref{AZX} and compared in Section~\ref{azalfacomp}.

\subsection{PCA metallicities}
\label{PCA}

One of the simplest approaches to estimate the metallicities of GCs is
the use of empirical metallicity calibrators, that is, photometric
and/or spectroscopic indicators which are sensitive to metallicity
effects and whose behaviors are well calibrated on the basis of GGCs
(Zinn \& West 1984; Armandroff \& Zinn 1988; Brodie \& Huchra 1991;
Carretta \& Gratton 1997; Kraft \& Ivans 2003). Certainly, the major
advantage of using metallicity calibrators is that the derived
metallicities are model independent. However, one has to deal with the
fact that the ages, abundance ratios and/or the metallicity range of
the GCs under study may not be the same as those in GGCs.  However,
given that the existence of young ($\la 5 $\,Gyr) GCs does not seem to
be commonplace in the GC systems of E galaxies (e.g.~Strader et
al.~2005), using metallicity calibrators on extragalactic GCs is a
reasonable starting point.

We have made use of the metallicity calibration by Strader \&
Brodie~(2004), which is based on a principal component analysis (PCA)
of 11 Lick indices for a sample of 39 GGCs. The first principal
component (PC1) is responsible for 94\% of the total variance, with
all the 11 indices having similar relative contributions. The fact
that the Balmer indices show opposite signs to the metal indices
suggests that PC1 is indeed correlated with the GGC metallicity. The
metallicity values ([m/H]) derived from PC1 for our 20 GCs are listed
in the second column of Table~\ref{AZX}, with uncertainties resulting
from the standard propagation of errors. As noted in Strader \&
Brodie, it is worth mentioning that [m/H], although being presumably
close to the Zinn \& West (1984) metallicity scale, may not strictly
reflect neither [Fe/H] nor overall metallicity ([Z/H]). In this sense,
a word of caution is reasonable before any comparison with the results
derived from other techniques. We refer the reader to Strader \&
Brodie~(2004) for a more detailed discussion on this topic.

Making use of the photometric data in F06 (see also
Table~\ref{basicdata}), Figure~\ref{PCA-color} illustrates the
corresponding color-metallicity relationship between the PCA
metallicities and F06 photometry.  The tight correlation existing
between both parameters supports the idea that the color distribution
of GC systems is mainly driven by metallicity. In agreement with
previous work on GC systems of other giant Es, we find the red GC
population to be very metal rich, with some GCs reaching [m/H] values
above solar. At the high metallicity regime ([m/H] $> -0.7$), the
steepening of the color-metallicity relationship qualitatively
resembles the one presented in Yoon, Yi \& Lee (2006) although, as
mentioned above, extrapolations beyond solar metallicity are not
completely safe and it could be just an artifact of the fact that the
PCA calibration relies on GGCs.

\begin{deluxetable}{lcr@{}c@{}cr@{}c@{}c}
\tablecolumns{8}
\tabletypesize{\scriptsize}
\tablewidth{0pt} \tablecaption{Age, metallicity and abundance
ratio of $\alpha$-elements ([$\alpha$/Fe]) derived for the GCs and the
central ($R_{\rm eff}$/8) region of NGC\,1407.\label{AZX} }
\tablehead{
\colhead{   } &          \multicolumn{1}{c}{PCA\tablenotemark{a}} &           \multicolumn{3}{c}{$\chi^2$\tablenotemark{b}}              &    \multicolumn{3}{c}{Mg, Fe and Balmer lines\tablenotemark{c}}     \\*
\colhead{ID } & \colhead{[m/H]}                   & \colhead{Age(Gyr)} & \colhead{[Z/H]} & \colhead{[$\alpha$/Fe]} & \colhead{Age(Gyr)} & \colhead{[Z/H]} & \colhead{[$\alpha$/Fe]}
}
\startdata
GC\#01 & $-0.81 \pm 0.03$ &$10.6 \pm 0.6$\ \ \ &$-0.78 \pm 0.03$\ \ \ &$+0.33 \pm 0.04$&$12.6 \pm 1.3$\ \ \ &$-0.85 \pm 0.06$\ \ \ &$+0.41 \pm 0.03$\\  
GC\#03 & $-0.22 \pm 0.07$ &$13.3 \pm 1.6$\ \ \ &$-0.33 \pm 0.04$\ \ \ &$+0.32 \pm 0.06$&$14.9 \pm 1.5$\ \ \ &$-0.36 \pm 0.03$\ \ \ &$+0.60 \pm 0.18$\\  
GC\#04 & $-0.69 \pm 0.05$ &$11.2 \pm 0.7$\ \ \ &$-0.65 \pm 0.04$\ \ \ &$+0.22 \pm 0.05$&$13.2 \pm 1.6$\ \ \ &$-0.70 \pm 0.07$\ \ \ &$+0.41 \pm 0.04$\\  
GC\#05 & $-0.50 \pm 0.06$ &$ 3.6 \pm 0.7$\ \ \ &$-0.08 \pm 0.06$\ \ \ &$+0.23 \pm 0.05$&$ 5.9 \pm 0.8$\ \ \ &$-0.26 \pm 0.07$\ \ \ &$+0.38 \pm 0.04$\\  
GC\#06 & $-0.78 \pm 0.05$ &$11.2 \pm 1.2$\ \ \ &$-0.93 \pm 0.06$\ \ \ &$+0.19 \pm 0.07$&$11.1 \pm 1.1$\ \ \ &$-0.91 \pm 0.04$\ \ \ &$+0.17 \pm 0.05$\\  
GC\#07 & $-0.41 \pm 0.07$ &$ 4.7 \pm 1.5$\ \ \ &$-0.20 \pm 0.14$\ \ \ &$+0.16 \pm 0.06$&$ 3.5 \pm 0.8$\ \ \ &$-0.21 \pm 0.09$\ \ \ &$+0.25 \pm 0.04$\\  
GC\#08 & $-1.01 \pm 0.09$ &$10.0 \pm 1.5$\ \ \ &$-1.03 \pm 0.08$\ \ \ &$+0.13 \pm 0.15$&$11.7 \pm 1.7$\ \ \ &$-1.10 \pm 0.07$\ \ \ &$+0.49 \pm 0.13$\\  
GC\#09 & $-1.44 \pm 0.06$ &$ 8.4 \pm 0.9$\ \ \ &$-1.35 \pm 0.06$\ \ \ &$-0.07 \pm 0.19$&$10.5 \pm 2.4$\ \ \ &$-1.57 \pm 0.15$\ \ \ &$-0.10 \pm 0.18$\\  
GC\#11 & $-0.27 \pm 0.10$ &$11.9 \pm 1.1$\ \ \ &$-0.43 \pm 0.05$\ \ \ &$+0.14 \pm 0.09$&$11.8 \pm 1.6$\ \ \ &$-0.43 \pm 0.07$\ \ \ &$+0.34 \pm 0.08$\\  
GC\#12 & $+0.03 \pm 0.10$ &$ 9.4 \pm 2.6$\ \ \ &$+0.00 \pm 0.09$\ \ \ &$+0.20 \pm 0.07$&$10.1 \pm 1.5$\ \ \ &$+0.02 \pm 0.07$\ \ \ &$+0.24 \pm 0.05$\\  
GC\#13 & $-0.82 \pm 0.08$ &$ 8.4 \pm 2.0$\ \ \ &$-0.70 \pm 0.12$\ \ \ &$+0.16 \pm 0.08$&$11.4 \pm 1.4$\ \ \ &$-0.83 \pm 0.06$\ \ \ &$+0.33 \pm 0.08$\\  
GC\#14 & $+0.12 \pm 0.08$ &$11.9 \pm 1.6$\ \ \ &$-0.18 \pm 0.05$\ \ \ &$+0.26 \pm 0.05$&$13.0 \pm 1.8$\ \ \ &$-0.12 \pm 0.04$\ \ \ &$+0.48 \pm 0.04$\\  
GC\#15 & $-0.31 \pm 0.09$ &$ 8.4 \pm 1.5$\ \ \ &$-0.23 \pm 0.06$\ \ \ &$+0.37 \pm 0.08$&$13.4 \pm 1.5$\ \ \ &$-0.33 \pm 0.04$\ \ \ &$+0.50 \pm 0.02$\\  
GC\#17 & $-0.08 \pm 0.07$ &$14.1 \pm 2.0$\ \ \ &$-0.18 \pm 0.05$\ \ \ &$+0.42 \pm 0.05$&$ 8.3 \pm 2.6$\ \ \ &$-0.15 \pm 0.07$\ \ \ &$+0.49 \pm 0.04$\\  
GC\#18 & $-0.29 \pm 0.10$ &$15.0 \pm 2.3$\ \ \ &$-0.35 \pm 0.07$\ \ \ &$+0.28 \pm 0.07$&$12.3 \pm 1.5$\ \ \ &$-0.30 \pm 0.05$\ \ \ &$+0.29 \pm 0.06$\\  
GC\#22 & $+0.35 \pm 0.10$ &$ 9.4 \pm 3.1$\ \ \ &$+0.15 \pm 0.12$\ \ \ &$+0.25 \pm 0.07$&$10.2 \pm 2.2$\ \ \ &$+0.06 \pm 0.10$\ \ \ &$+0.26 \pm 0.05$\\  
GC\#26 & $-1.10 \pm 0.07$ &$11.2 \pm 1.1$\ \ \ &$-1.05 \pm 0.05$\ \ \ &$+0.41 \pm 0.11$&$11.4 \pm 2.1$\ \ \ &$-1.21 \pm 0.08$\ \ \ &$+0.50 \pm 0.12$\\  
GC\#27 & $+0.24 \pm 0.11$ &$12.6 \pm 2.0$\ \ \ &$+0.00 \pm 0.06$\ \ \ &$+0.32 \pm 0.07$&$12.7 \pm 1.6$\ \ \ &$+0.01 \pm 0.06$\ \ \ &$+0.60 \pm 0.13$\\  
GC\#30 & $-0.12 \pm 0.18$ &$ 7.9 \pm 3.3$\ \ \ &$-0.20 \pm 0.13$\ \ \ &$+0.29 \pm 0.13$&$ 9.7 \pm 2.1$\ \ \ &$-0.14 \pm 0.07$\ \ \ &$+0.52 \pm 0.09$\\  
GC\#33 & $-0.96 \pm 0.13$ &$ 3.2 \pm 1.8$\ \ \ &$-0.55 \pm 0.28$\ \ \ &$+0.26 \pm 0.19$&$ 5.2 \pm 1.7$\ \ \ &$-0.82 \pm 0.14$\ \ \ &$+0.40 \pm 0.14$\\  
R$_{\rm eff}$/8& \nodata  &$11.9 \pm 1.4$\ \ \ &$+0.38 \pm 0.04$\ \ \ &$+0.33 \pm 0.01$&$12.0 \pm 0.4$\ \ \ &$+0.33 \pm 0.03$\ \ \ &$+0.34 \pm 0.01$\\  
\enddata
\tablecomments{Values derived using different techniques: (a)
Metallicity calibration based on a principal component analysis of
Lick indices (Strader \& Brodie 2004); (b) Multivariate analysis of
Lick indices (Proctor et al.~2004); (c) Weighted combination of
solutions driven from single index-index diagrams based on Mg$b$,
$<$Fe$>$, [MgFe]' and the Balmer Lick indices}
\end{deluxetable}

\subsection{Ages, metallicities and abundance ratios from SSP models}
\label{index-index}

\subsubsection{Using a multivariate analysis}
\label{chi2}

Metallicities ([Z/H]), ages and [$\alpha$/Fe] ratios for the GCs and
galaxy starlight were computed from their Lick indices by performing
multivariate fits to the TMB03 and TMK04 SSP models. For comparisons,
the same procedure was performed for a sample of 41 GGCs by Schiavon
et al.~(2005). The application of this technique to GCs is described
in detail in Proctor et al.~(2004).  Briefly, 12--18 indices were fit
simultaneously for [Z/H], age and [$\alpha$/Fe] to the grid of models
until $\chi^2$ was minimized. Any indices which deviated from the fit
were excluded and $\chi^2$ recalculated (here, we employed three
iterations with rejections of 5$\sigma$--3$\sigma$--5$\sigma$). It is
important to keep in mind that this method makes use of all the
available indices, so the $best$ solution relies on much more
information than if it had been derived from a single, index-index
diagram. This advantage may turn into a problem because of the fact
that the behavior of some elements (like, e.g.~C, N and Ca) is not
well understood and/or accounted for the models, so the inclusion of
their corresponding indices --or indices affected to some extent by
these elements-- in the $\chi^2$ procedure may affect the reliability
of the final result. For this reason, several Lick indices were
routinely rejected during this process: i) CN$_{1}$ and CN$_{2}$; as
it will be discussed later, these indices are extremely enhanced
relative to the models; ii) Ca4227; it exhibits a large scatter around
the models without any reasonable explanation. Actually, despite Ca is
an alpha element like Mg, its behavior is not well understood yet (see
Worthey 1992, Vazdekis et al.~1997, Trager et al.~1998, Saglia et
al.~2002, Cenarro et al.~2003, Thomas et al.~2003b, Cenarro et
al.~2004) and we preferred to exclude it from the present analysis;
iii) Fe4531 and Fe5015; as they look systematically wrong --unlike
other Fe indices-- in the GGC comparison sample. Errors in the output
parameters were computed by means of Montecarlo simulations according
to the individual errors of all the indices involved.

Acording to this method, Figure~\ref{fighistog} represents the ages,
metallicities and [$\alpha$/Fe] ratios derived for our 20 GCs (filled
circles), the $R_{\rm eff}/8$ central spectrum of NGC\,1407 (filled
star), and the sample of 41 GGC spectra (open squares) from Schiavon
et al.~(2005). Histograms of the three parameters are provided for
both GC samples.

The parameters inferred for GGCs allow us to constrain the reliability
of the method. In the age-[Z/H] plane of Fig.~\ref{fighistog}, MP GGCs
tend to be younger than MR ones, whereas no such an age-metallicity
relationship is known to exist on the basis of accurate CMDs (e.g.~de
Angeli et al.~2005). Such a fictitious trend might be an artifact of
the SSP models that, accounting for BHB effects for [Z/H] $\leq -0.5$,
make the Balmer line strengths to turn up for ages older than $\sim
9$\,Gyr. The old-age regimes of Balmer index-index diagrams thus
become bi-valuated and objects with low Balmer indices that lie below
the model grids are systematically assigned to be younger than if they
were interpreted on the basis of non-BHB effect models. Probably, this
inconsistency is in turn driving the extremely large [$\alpha$/Fe]
values obtained for GGCs at the low metallicity end. Because of the
above limitation, and given that typical uncertainties in dating old
stellar populations are in general very large, trying to distinguish
small differences among the ages of {\sl old} GCs in NGC\,1407 is
beyond the scope of this paper. Fortunately, our GC set does not reach
[Z/H] values as low as the GGCs and the derived [$\alpha$/Fe] values
are still expected to be reasonable.

Keeping in mind the systematic trends introduced by the procedure, it
is clear from Fig.~\ref{fighistog} that most GCs in NGC\,1407 are
found to be old like GGCs ($\sim 11$\,Gyr). However 3 of them (GC\#05,
\#07 and \#33, see Table~\ref{AZX}) formally have ages of $\sim
4$\,Gyr. We defer to Section~\ref{YBHB} a discussion on the existence
of these potentially young GCs (hereafter {\sl y}GCs) in NGC\,1407.
For the old GCs only, we compute a mean age of $10.9 \pm 0.5$\,Gyr and
a {\sl rms} standard deviation of 2.1\,Gyr. Many of them follow the
same, fictitious age-metallicity trend as GGCs, although a few GCs
with large error bars look slightly younger. In agreement with the
metallicities derived according to the PCA method, we find that the
GCs in NGC\,1407 range from [Z/H]$\sim -1.5$\,dex to slightly above
solar. For the [$\alpha$/Fe] ratios, we find most GCs have
$\alpha$-enhanced chemical compositions. In particular, the mean
[$\alpha$/Fe] ratio for the whole sample is $+0.24 \pm 0.03$\,dex with
a {\sl rms} standard deviation of 0.11\,dex.  Unlike Puzia et
al.~(2005), we do not detect any significant correlation between
[$\alpha$/Fe] and any of the other two parameters. Finally, the
integrated, $R_{\rm eff}$/8 central spectrum of NGC\,1407 is found to
be old ($11.9 \pm 1.4$\,Gyr), metal-rich ([Z/H] $= +0.38 \pm
0.04$\,dex) and $\alpha$-enhanced ([$\alpha$/Fe] $= +0.33 \pm
0.01$\,dex).

\subsubsection{Using an iterative procedure based on Mg, Fe and Balmer indices}
\label{iterative}

In addition to the above procedure, we carried out an iterative method
to determine the ages, metallicities and [$\alpha$/Fe] ratios of our
GCs and galaxy spectra according to index-index diagnostics based on
the Balmer Lick indices (H$\delta_{A}$, H$\delta_{F}$, H$\gamma_{A}$,
H$\gamma_{F}$ and H$\beta$) as well as [MgFe]', Mg$b$ and $<$Fe$>$. A
similar procedure is carried out in Puzia et al.~(2005).

Figure~\ref{index-index-fig1} illustrates the
age-metallicity-[$\alpha$/Fe] diagnostic diagrams constructed on the
basis of the above indices. To follow the location of the three {\sl
y}GCs inferred in Section~\ref{chi2}, different symbols are assigned:
circles (old GCs) and triangles ({\sl y}GCs). In turn, an additional
color code is employed to distinguish blue GCs ($B - I < 1.87$;
black-filled symbols) from red GCs ($B - I > 1.87$; open and
grey-filled symbols, see below). The star corresponds to the $R_{\rm
eff}/8$ central spectrum of NGC\,1407 and, for comparisons, asterisks
represent the sample of 41 GGCs of Schiavon et al.~(2005).

In panel $a$, model predictions for Mg$b$ and $<$Fe$>$ at fixed age
(11\,Gyr) and varying metallicity (dashed lines) and [$\alpha$/Fe]
ratios (different-width solid lines) are over-plotted. In particular,
red GCs with high [$\alpha$/Fe] ratios ($ > 0.4$\,dex; GC\#03, \#14,
\#15, \#17, \#27 and \#30; see last column in Table~\ref{AZX}) are
displayed as open circles to allow following their locations
throughout different diagnostic diagrams. Hereafter, we will refer to
them as $\alpha$GCs. The rest of GCs roughly reproduce typical [Mg/Fe]
values of GGCs. In particular, we will refer as {\sl non}-$\alpha$GCs
to those red GCs with typical GGC [Mg/Fe] ratios (grey-filled
symbols).

Panels $b - f$ display the model predictions with constant
[$\alpha$/Fe] ratio ($= +0.3$) and varying age (dotted lines) and
metallicity for the Balmer Lick indices and [MgFe]'. The latter is
employed as metallicity indicator since, on the basis of TMB03 models,
it is defined to be insensitive to different [$\alpha$/Fe] ratios. To
guide the eye, the 15\,Gyr line is represented as a solid line and the
solar-metallicity one as dash-dotted line. It is clear from the figure
that the GCs in NGC\,1407 basically follow an old-aged, metallicity
sequence and exhibit [$\alpha$/Fe] ratios larger than solar. Also, the
Balmer indices of {\sl y}GCs are systematically larger than those of
other GCs with similar [MgFe]' values --especially in the H$\delta$
and H$\gamma$ diagrams (panels $b - e$)--, thus providing support to
the age differences reported in Section~\ref{chi2}.

The ultimate aim of this method is to find, for each object, the best
solution of age, metallicity and [$\alpha$/Fe] inferred on the basis
of two diagnostic diagrams: i) the plane Mg$b$ vs $<$Fe$>$ (panel
$a$), and ii) the plane H$_{\chi}$--[Mg/Fe]', where H$_{\chi}$ refers
to one of the five possible Balmer line indices (panels $b$, $c$, $d$,
$e$ and $f$). In this sense, by performing this technique for each
Balmer line index, five different solutions of age, metallicity and
[$\alpha$/Fe] are obtained.  In particular, for a certain Balmer
index, the iterative procedure carried out to infer the final
solutions can be summarized in the following diagnostic sequence: 1)
Mg$b$--$<$Fe$>$ (age$_0$); 2) H$_{\chi}$--[Mg/Fe]'
([$\alpha$/Fe]$_1$); 3) Mg$b$--$<$Fe$>$ ([Z/H]$_2$); 4)
H$_{\chi}$--[Mg/Fe]' (age$_3$); 5) Mg$b$--$<$Fe$>$ ([Z/H]$_4$); and 6)
H$_{\chi}$--[Mg/Fe]' ([$\alpha$/Fe]$_5$). For each diagnostic diagram
of the sequence, the models are displayed at a fixed parameter
--quoted within parenthesis-- that corresponds, as indicated in the
subscript, to the solution derived in the previous step. An age$_0$ of
11\,Gyr was assumed as starting point of the sequence. The plane
Mg$b$--$<$Fe$>$ at fixed [$\alpha$/Fe] was deliberately avoided within
the sequence given that it is completely degenerate. Instead, steps 3
and 2 are respectively repeated in 5 and 6 to close the cycle. No grid
extrapolations were applied. Instead, for objects lying outside the
models, the closest\footnote{The minimum distance between a given
point and the model grids has been computed taken into account the
errors of the indices, that is, by using a re-scaled space in which
errors in both axes have the same size.} point of the grid was
considered as the most reliable solution. In all cases, the method
converged after two iterations. If no consistent solutions between the
two panels could be found, means of the two different determinations
of age, [Z/H] and [$\alpha$/Fe] were computed as final values.

For each object, errors in the ages, metallicities and [$\alpha$/Fe]
ratios were estimated according to its index errors by performing
MonteCarlo simulations on the model grids. In turn, large extrapolated
uncertainties were assigned to those objects falling outside the
models.  The final ages, metallicities and [$\alpha$/Fe] ratios
obtained for each object are listed in Table~\ref{AZX} and correspond
to error-weighted means of the five solutions derived from the five
different sequences. The final errors in this table were computed as
the larger of either the typical error of the error-weighted mean or
the error-weighted standard deviation of the five single
determinations.

Concerning the $R_{\rm eff}$/8 central spectrum of NGC\,1407, the
current method confirms the results given in the previous section that
the galaxy is old, metal rich and $\alpha$-enhanced (see values in
Table~\ref{AZX}). The ages derived here for the galaxy are at odds
with the $\sim 2$\,Gyr reported by Denicol\'o et al.~(2005) on the
basis of H$\beta$. The reasons for such a discrepancy seem to arise
from the emission corrections of H$\beta$ computed in that paper
since, as pointed out by the authors, they are systematically larger
than the ones presented in previous work.  We do not find any
significant evidence for emission filling H$\beta$, so the index value
we employ in this paper is not corrected for emission. Interestingly,
the H$\delta$ and H$\gamma$ Lick indices we have measured for the
galaxy are fully consistent (within the errors) with the ones given
by Denicol\'o et al., but they did not use them for age estimations.
Finally we note that Thomas et al.~(2005) have also published a
central age for NGC\,1407 using the same SSP models as here and
estimate a relatively old age of $7.4 \pm 1.8$\,Gyr. On the basis of
surface brightness fluctuations, Cantiello et al.~(2005) also reported
an age of $\sim 11$\,Gyr for NGC\,1407.

\subsection{Age, metallicity and [$\alpha$/Fe] ratio comparisons}
\label{azalfacomp}

Figure~\ref{figcomp4} illustrates a comparison between the ages,
metallicities and [$\alpha$/Fe] ratios derived from the different
approaches performed in Sections~\ref{PCA} and \ref{index-index}

In panel $a$, the metallicities derived from the PCA calibration are
compared with those obtained from the $\chi^2$ procedure. For old GCs
(filled circles), there exists a total agreement in the low
metallicity regime that progressively turns into a small offset ($\sim
0.12 \pm 0.03$\,dex) for metallicities larger than $-0.5$\,dex. As
mentioned in Section~\ref{PCA}, we suspect that the PCA metallicities
are overestimated in this regime because of the lack of very
metal-rich (MR) GGCs constraining the metallicity
calibration. Differences in the mean [$\alpha$/Fe] values of both GC
sets could also be driving the observed offset, although issues
concerning the definition of the metallicity scales should not be
ruled out either. Finally, the three {\sl y}GCs (open triangles)
systematically deviate from the 1:1 relationship as a consequence of
the age-metallicity degeneracy. If, as we will discuss later, {\sl
y}GCs were old GCs hosting BHBs, the PCA metallicities derived for
these clusters would be a better estimation of their true metal
contents.

In panels $b$, $c$ and $d$, the ages, metallicities and [$\alpha$/Fe]
ratios derived from the iterative method based on Mg, Fe and Balmer
Lick indices are respectively compared with the ones inferred from the
$\chi^2$ procedure. Once again, the agreement in metallicity is very
good. With a {\sl rms} standard deviation of $\sim 0.08$\,dex, the
Mg-Fe-Balmer diagnostic provides slightly lower values in metallicity
(by $\sim 0.04 \pm 0.02$\,dex; see panel $b$), which is a consequence
of both the age-metallicity degeneracy and the fact that the ages are,
in most cases, found to be slightly older (by $\sim 0.8 \pm 0.6$\,Gyr;
panel $c$). Therefore, on the basis of the Mg-Fe-Balmer diagostic, a
mean age of $11.7 \pm 0.4$\,Gyr is obtained for the old GC population,
with a {\sl rms} standard deviation of $\sim 1.6$\,Gyr. Interestingly,
with ages in the range $\sim 3 - 6$\,Gyr, {\sl y}GCs are again the
{\it youngest} GCs of the sample. The [$\alpha$/Fe] ratios provided by
the Mg-Fe-Balmer procedure are essentially the ones inferred from the
Mg$b$-$<$Fe$>$ plane so, unlike the $\chi^2$ techique, no other
elements/indices constrain the final values. The mean [$\alpha$/Fe]
ratio for the whole GC sample is $0.38 \pm 0.04$\,dex (with a {\sl
rms} standard deviation of 0.18\,dex), that is, $\sim 0.14$\,dex
larger than the one derived from the $\chi^2$ procedure (panel $d$).
The offsets in the mean ages and [$\alpha$/Fe] ratios derived from
both techniques are thus readily explained if we consider that
H$\delta$ and H$\gamma$ indices are predicted to increase with the
increasing $\alpha$-enhancement (TMK04).

It is worth noting that systematic differences among the values in
panels $b$, $c$ and $d$ are just due to the distinct techniques used
to infer ages, metallicities and [$\alpha$/Fe] ratios, as both the
data and the SSP model predictions are the same in both cases. In this
sense, it should be considered as an additional uncertainty that
deserves to be taken into account, as age differences of a few
Gyr may suffice to support or rule out different GC formation
scenarios.

\section{C and N-sensitive Lick indices}
\label{CN}

Having discussed the behavior of $\alpha$-elements as a whole, in this
section we explore, from a qualitative point of view, the behavior of
C and/or N-sensitive Lick indices in our GC set. With this aim,
Figure~\ref{index-index-fig3} illustrates some interesting diagnostic
diagrams, with symbols and models being the same as those presented in
Fig.~\ref{index-index-fig1}. Note that, since the models we are using
basically account for enhancement of $\alpha$-elements at constant C
and N, their predictions must be interpreted in a relative way rather
than to constrain C and N absolute abundances.

In panels $a$ and $b$ we find clear correlations --presumably driven
by metallicity-- between CN$_2$ (measuring the strength of the
cyanogen molecule; CN) and the indices [MgFe]' and $<$Fe$>$. What is
even more remarkable is the striking evidence for CN enhancement in
NGC\,1407 GCs over the full metallicity range.  In particular, at
fixed [MgFe]' values, CN$_2$ indices in NGC\,1407 GCs are
systematically larger than those of GGCs, with index differences
increasing from $\Delta$CN$_2 \sim 0.04$\,mag (for MP GCs) up to
$\Delta$CN$_2 \sim 0.09$\,mag (MR GCs). In fact we note that some MR
GCs in NGC\,1407 reach --and even go beyond-- the CN$_{2}$ values of
the galaxy, which, in turn, exhibits a very strong CN$_2$ index within
$R_{\rm eff}/8$ as expected for very massive Es (S\'anchez-Bl\'azquez
et al.~2003). It is important to note here that the error and {\sl
rms} of our CN$_2$ correction to the Lick system (see
Table~\ref{Lickset}) are small enough that they should not be
responsible for any of the above results.

As reported by Tripicco \& Bell (1995), the Fe4668 Lick index
(hereafter C$_2$4668) is extremely sensitive to carbon abundance. In
fact, as it goes as the square of the C abundance, C$_2$4668 rises
steeply with increasing metallicity and the C$_2$ bands may be clear
for the most metal-rich systems, with the bands being dominated by
other species at the very low metallicity regime. As compared to
CN$_2$, C$_2$4668 values in NGC\,1407 GCs exhibit a quite different
behavior (panel $c$). At first sight, it is clear that they are not
systematically larger than the model predictions. In addition, both
NGC\,1407 GCs and GGCs follow a {\it similar} C$_2$4668--$<$Fe$>$
sequence: at the low-metallicity regime they both are essentially
consistent with the models, whilst a larger C$_2$4668 spread is
apparent for MR GCs in NGC\,1407. Also, unlike CN$_2$, C$_2$4668 is
clearly stronger in the galaxy than in their GCs (even in the MR
ones). Interestingly, the scatter in C$_2$4668 exhibited by MR GCs is
not random: $\alpha$GCs (open circles) also possess larger C$_2$4668
values. Such a difference in the C$_2$4668--$<$Fe$>$ diagram cannot be
explained solely by age effects, since it would require age
differences of order a Hubble time, inconsistent with the old GC ages
derived on the basis of other Lick indices (Sections~\ref{chi2} and
\ref{iterative}).  Therefore, the most plausible interpretation is
that C abundances are driving the observed differences: apart from
larger [Mg/Fe] ratios, $\alpha$GCs possess [C/Fe] ratios larger than {\sl
non}-$\alpha$GCs. Moreover, given that i) the location of $\alpha$GCs
in the C$_2$4668--$<$Fe$>$ diagram is similar to the one exhibited by
MR GGCs, and ii) {\sl non}-$\alpha$GCs (grey-filled symbols) lie
systematically below the models, one should argue for the existence of
C underabundances for {\sl non}-$\alpha$GCs.

In panel $d$, the C-sensitive (CH) G4300 band is plotted versus
$<$Fe$>$. Once again, $\alpha$GCs --except for the one with the
largest errors-- exhibit mean G4300 values larger than {\sl
non}-$\alpha$GCs which basically trace the locus of GGCs. This
behavior provides independent support to the above hypothesis that C
is either particularly enhanced in $\alpha$GCs or depressed in {\sl
non}-$\alpha$GCs. As an obvious consequence, the G4300--C$_2$4668
plane in panel $e$ exhibits $\alpha$GCs populating the high
metallicity regime of both C-sensitive indices.

Finally, in order to compare the behaviors of N and C-sensitive
indices, CN$_{2}$ and C$_2$4668 are represented in panel $f$. MP GCs
distribute along a linear CN$_{2}$-C$_2$4668 sequence with a slope
similar to the one predicted by the models, even though there exists a
clear offset presumably due to the high CN$_{2}$ values previously
reported. The picture is different for MR GCs, as they strongly
deviate from the trend predicted by the models. In particular,
CN$_{2}$ in {\sl non}-$\alpha$GCs increases dramatically as C$_2$4668
has basically saturated (C underabundances). On the other hand,
$\alpha$GCs lie systematically closer to the model predictions because
of their higher C abundances, although they are still far from
following the sequence drawn by MP GCs.

To summarize, we report extremely strong CN bands in the GCs of
NGC\,1407. Interestingly, we find unprecedented hints for the
existence of two MR GC regimes with different abundance patterns for
Mg and C. Plausible interpretations on the origin of these
subpopulations will be discussed in the next section.

\section{Discussion}
\label{discussion}

As a result of the analysis and interpretation of Lick indices carried
out in the previous section, we conclude that the GC system of
NGC\,1407 basically follows an old-aged, metallicity sequence,
reaching metallicities above solar and exhibiting an enhancement of
$\alpha$-elements. Overall, these results are in reasonable agreement
with previous studies on GC systems of Es (e.g.~Puzia et
al.~2005). Interestingly, we have also found hints for the existence
of younger GCs (the so-called, {\sl y}GCs), as well as evidence for
two different abundance patterns among MR GCs in NGC\,1407
($\alpha$GCs vs {\sl non}-$\alpha$GCs). We devote the present section
to discuss the origin and reliability of these results.

\subsection{Young GCs or blue horizontal branch effects?}
\label{YBHB}

Although the origin of the ``second parameter'' of GGCs (Sandage \&
Wallerstein 1960; van den Bergh 1967; Sandage \& Wildey 1967) is still
a controversial issue (e.g.~Lee et al.~1994), there exists a well
established consensus that the existence of blue horizontal branch
(BHB) morphologies is driven by high ratios between the mass of the He
burning core and the total stellar mass. Therefore, mass-loss along
the red giant branch (RGB) and/or high He abundances are ideal
candidates to explain the observed BHB morphologies. Because of the
contribution of luminous, A-type stars with $T_{\rm eff} \sim
10000$\,K, the integrated spectra of old GCs having BHBs exhibit
strong Balmer lines (de Freitas Pacheco \& Barbuy 1995; Lee et
al.~2000; Maraston \& Thomas 2000; Maraston et al.~2003; Schiavon et
al.~2004, hereafter SCH04) that may be wrongly interpreted in terms of
younger ages. Note also that, although it is not discussed in this
paper, a similar effect could arise from the presence of a
considerable population of blue straggler stars (e.g Rose 1985; Rose
1994; Xing \& Deng~2005). Hence, could it be that the {\sl y}GCs
(\#05, \#07 and \#33) reported in Section~\ref{index-index} are,
actually, old GCs with BHBs?

\subsubsection{Metal-poor GCs}

In order to address the above question, we first aim to carry out a
more detailed comparison between our data and the model predictions
used in this paper. Based on the HB morphologies observed in GGCs, the
models by TMB03 and TMK04 include BHB effects for metallicities below
$-0.5$. Bearing this upper limit in mind, and provided the
metallicities given in Table~\ref{AZX} for {\sl y}GCs, only GC \#33
can be considered well within the metallicity regime in which BHB
effects are being accounted for in the models. Note also that, if a
BHB were indeed responsible for the high Balmer lines of this cluster,
the metallicities estimated in Section~\ref{index-index} would have
been overestimated due to the implicit age-metallicity degeneracy of
the grids, so the {\sl true} metallicity would still be safe for the
sake of comparison. The PCA metallicity could thus be considered as a
more reasonable value for this particular case.

In Fig.~\ref{index-index-fig1} (panels $b - f$), BHB effects are
apparent in the model grids as the varying metallicity lines at fixed
15\,Gyr (solid lines) lie above the predictions for slightly younger
SSPs. Within the $1\sigma$ error bars, GC \#33 (black-filled triangle)
is consistent with the solid line of each diagram except for
H$\delta_{A}$, although even in this case it is certainly close to
that line as well. The fact that most GCs in
Fig.~\ref{index-index-fig1}$c$ lie systematically below the models
might suggest that a systematic offset in the H$\delta_{F}$ data is
artificially reconciling GC \#33 with the BHB predictions. However,
the excellent agreement between NGC\,1407 GCs and GGCs would indeed be
difficult to interpret if such an offset existed. In any case, it
seems clear that BHB effects in old stellar populations, as calibrated
in TMB03 and TMK04, are able to reproduce the strengths of H$\beta$,
H$\gamma$ and, probably, H$\delta$ in GC \#33.

\begin{deluxetable}{lcccccc}
\tablecolumns{6}
\tabletypesize{\scriptsize}
\tablewidth{0pt} \tablecaption{Offsets in the Balmer Lick indices of MR stellar populations due to BHB effects. \label{BHBtest}}
\tablehead{
\colhead{          } & \colhead{$\Delta$H$\delta_{A}$ (\AA)} & \colhead{$\Delta$H$\delta_{F}$ (\AA)} &\colhead{$\Delta$H$\gamma_{A}$ (\AA)} &\colhead{$\Delta$H$\gamma_{F}$ (\AA)} &\colhead{$\Delta$H$\beta$ (\AA)}  \\
}
\startdata
MR GGCs + BHB (Puzia et al.~2004)\tablenotemark{a}              & 2.0 & 1.0 & 3.3 & 1.4 & 0.4 \\
${\rm [Z/H]}_{\sun}$ SSP + BHB (Maraston 2005)\tablenotemark{b} & 4.8 & 5.7 & 2.4 & 2.9 & 0.9 \\

MR {\sl y}GCs in NGC\,1407\tablenotemark{c} & $2.15 \pm 0.60$ & $1.41 \pm 0.45$ & $1.80 \pm 0.49$ & $1.69 \pm 0.37$ & $0.47 \pm 0.28$ \\
\enddata
\tablecomments{
Offsets in the Balmer Lick indices of: (a) metal-rich GGCs, with
offsets arising from the existence of BHB morphologies (Puzia et
al.~2004); (b) solar-metallicity, SSP models by Maraston (2005), with
offsets accounting for BHB effects; and (c) metal-rich, presumably
{\it young} GCs in NGC\,1407 (\#05 and \#07) as compared to a
subsample of metal-rich, old GCs in this galaxy with similar [MgFe]'
values (\#03, \#11, \#15 and \#18). The offsets in (c) are comparable
to and lower than those in (a) and (b) respectively, so BHB effects
cannot be ruled out to explain the large Balmer Lick indices
in (c). See more details in the text.}
\end{deluxetable}

\subsubsection{Metal-rich GCs}

As reported by Greggio \& Renzini (1990), BHBs are not just reserved
for MP stellar populations but they are also possible in GCs with
metallicities above solar. Based on a comparison between MR GGCs with
BHBs (NGC\,6388 and NGC\,6441) and those with similar metallicities
hosting just a red HB (RHB; NGC\,6356 and NGC\,6637), Puzia et
al.~(2004) set limits to the effect of BHBs on the five Balmer Lick
indices. As pointed out by the authors, such an empirical comparison
was just made on a local frame and the existence of more extreme BHB
morphologies in extragalactic GCs cannot be ruled out.  In order to
check whether the low ages derived for the two MR, {\sl y}GCs (\#05
and \#07; grey-filled triangles in Fig.~\ref{index-index-fig1}) may be
the by-product of BHB morphologies, we have carried out a similar
comparison. Unfortunately, we do not have information about the HB
morphologies of the GCs in NGC\,1407, so it is formally impossible
defining a reference sample of GCs with RHBs. Instead, since [Mg/Fe]'
is not affected by BHB effects (Puzia et al.~2004), we have
established our reference sample to be all those GCs that i) have the
same [Mg/Fe]' values (within the errors) than GCs \#05 and \#07 ($\sim
2.45$\,\AA; see Table~\ref{Lickindices}) --thus ensuring the effect of
metallicity on each single Balmer line-strength to be the same in both
subsamples of GCs--, and ii) have been separately confirmed to be old
by the two different approaches described in
Section~\ref{index-index}. The GCs fulfilling the above requirements
are \#03, \#11, \#15 and \#18, the ones will be referred to as {\sl
non-BHB}\,GCs.

In order to consider the most extreme cases, for each Balmer line we
have computed the difference between the highest index value of {\sl
y}GCs and the lowest index value of {\sl non-BHB}\,GCs.  The offsets
derived in this way are shown in Table~\ref{BHBtest} together with the
ones computed by Puzia et al.~(2004) for GGCs. In all cases, the
Balmer index offsets between {\sl y}GCs and {\sl non-BHB}\,GCs in
NGC\,1407 are consistent with the ones existing between BHB and RHB,
MR GGCs.

New SSP models accounting for BHB effects in the high-metallicity
regime have been recently presented by Maraston (2005). For
illustration, the offsets corresponding to a solar metallicity SSP
--as predicted by that work-- are given in the second row of
Table~\ref{BHBtest}. In spite of the fact that Maraston quotes the BHB
effect on H$\beta$ as a relative variation, we have transformed it
into an absolute, index offset by assuming the typical H$\beta$ value
of a solar metallicity, 15\,Gyr old SSP with [$\alpha$/Fe] = $+
0.3$. Once again, it is clear that they are significantly larger than
the ones we measure for our MR, {\sl y}GCs, thus suggesting that the
existence of BHBs in GCs \#05 and \#07 cannot be ruled out either.

\subsubsection{BHB spectral diagnostic}

To conclude the present discussion, we make use of the method proposed
by SCH04 to identify BHBs in the integrated spectra of GCs. In short,
if H$\delta$, H$\beta$ and any Fe line are measured accurately, the
presence of BHB stars in a given GC can be inferred if the age derived
from the ratio H$\delta$/ H$\beta$ is substantially younger than the
one indicated just by H$\beta$. Basically, the above recipe relies on
the fact that BHB stars --similar to blue straggler stars-- constitute
an extra, hot component in old stellar population that essentially
contributes at short wavelengths. In this sense, the relative
contribution of BHB stars to the integrated H$\delta$ is larger than
it is for H$\beta$. In other words, if one uses SSP model predictions
that {\it do not} account for BHB effects --or even if they do it but
it is underestimated--, the ages derived from H$\delta$ for GCs with
BHBs should be younger than the ones derived using H$\beta$, whilst
should be the same for GCs with RHBs. The last argument also holds for
MR GCs, as it is proven in SCH04 for NGC\,6388 and NGC\,6441 (MR GGCs
with BHBs). It is clear however that this technique is model-dependent
and different stellar population models may drive different solutions.

Interestingly, the integrated Balmer lines of our {\sl y}GCs exhibit
qualitatively the same trend as predicted by SCH04 for GCs with BHB
stars. At first sight, it is clear from Fig.~\ref{index-index-fig1}$b
- f$ that the relative differences among the Balmer indices of {\it
y}GCs and old GCs with similar [MgFe]' values are larger for H$\delta$
and H$\gamma$ than for H$\beta$. Specifically, the individual
ages derived from H$\beta$ and H$\delta_{A}$
\footnote{We employ H$\delta_{A}$ rather than H$\delta_{F}$ since, as
mentioned above, the model predictions for H$\delta_{F}$ seem to be
overestimated as compared to GGCs and our GC set, thus deriving
unreliable, extremely old ages in most cases} as result of the
iterative procedure described in Section~\ref{index-index} are,
respectively, $10.8 \pm 2.8$ and $5.1 \pm 1.3$\,Gyr (\#05), $9.2 \pm
2.9$ and $6.6 \pm 2.6$\,Gyr (\#07), and $6.0 \pm 3.8$ and $2.8 \pm
1.6$\,Gyr (\#33), thus confirming the relative age differences
predicted by SCH04. As an additional test, Figure~\ref{figBHB}
illustrates the H$\delta_{A}$/H$\beta$ versus $<$Fe$>$ diagnostic
diagram proposed by SCH04 to confirm BHB effects. It is clear that our
{\sl y}GCs lie systematically above the cloud of old GCs of the same
metallicity. Using for each cluster its corresponding [$\alpha$/Fe]
ratio given in Table~\ref{AZX}, the ages derived from
Figure~\ref{figBHB} are $\sim 3.5$\,Gyr (\#05), $\sim 6.0$\,Gyr
(\#07), and $\sim 1.0$\,Gyr (\#33). Summarizing, the sequence of ages
derived for our {\sl y}GCs on the basis of H$\beta$, H$\delta_{A}$ and
H$\delta_{A}$/H$\beta$ follow a decreasing trend as predicted by SCH04
for SSPs that host BHB stars.

We therefore conclude that, although we cannot completely reject the
possibility that our {\sl y}GCs are indeed younger than the rest of
the GC sample, our results for the {\sl y}GCs are well explained by
both BHB model predictions and BHB line-strength diagnostics. In
addition we note that, as compared to genuine young GCs populations
found in merger remnant Es (e.g.~NGC\,3610; Strader et al.~2004), we
find the behavior, location and distribution of our {\sl y}GCs to be
quite different. The young GCs found by Strader et al. fall around the
MR, young regime of typical index-index, age-metallicity, diagnostic
diagrams, thus clearly deviating from the locus of the old, {\sl
normal} GCs. Instead, our {\sl y}GCs span over a wider range of
metallicity --in fact, they do not belong to the MR end of GCs in
NGC\,1407-- and they just lie slightly above the sequence of old GCs
in panels $b - f$ of Fig.~\ref{index-index-fig1}. In addition, unlike
the merger remnant NGC\,3610 ($2 - 3$ Gyr old), the existence of young
GCs would be difficult to reconcile with the luminosity-weighted age
of $\sim 10 - 12$\,Gyr we find for the R$_{\rm eff}$/8 central
spectrum NGC\,1407.

\subsection{Element abundance ratios in GCs}
\label{CNO}

In Section~\ref{CN}, we have reported, for the first time, evidence
for two distinct chemical compositions among MR GCs in an E galaxy,
i.e. a subset of MR GCs have larger [Mg/Fe] {\it and} [C/Fe]
ratios. Here we discuss possible interpretations for the observed
differences on the basis of current theories of chemical evolution.

\subsubsection{GC formation time-scales}

Since different elements are preferentially produced by stars of
different lifetimes, element abundance patterns are essential to
constrain the mechanisms and time-scales of star formation in stellar
populations. For instance, the fact that $\alpha$-elements (e.g.~Mg,
Si, Ti) are rapidly ejected to the interstellar medium by massive,
Type II SNe (Faber et al.~1992; Worthey et al.~1992; Matteucci 1994),
whereas Fe is produced in Type Ia SNe a few gigayears later, means
high [$\alpha$/Fe] ratios are often interpreted as a fingerprint of
rapid star formation events. Elements like C and N, however, can be
produced in stars over a wide range of masses by means of different
mechanisms (see e.g.~Kobayashi et al.~2006). C is a primary element
(that is, an element formed by stellar, nuclear fusion) which can be
produced by both massive and intermediate-mass stars. Because of
uncertainties in the C yield modeling, it is still a matter of debate
whether C is mainly produced in massive stars (Carigi 2000; Henry et
al.~2000) or in low and intermediate-mass stars (Chiappini et
al.~2003). N is known to have a double nature. It is basically a
secondary element (an element that requires a previous seed to be
produced) as it is a by-product of the CNO cycle formed at the expense
of pre-existing C. In turn, N is also a primary element that may be
formed in intermediate-mass stars during the third dredge-up of the
AGB (Renzini \& Voli 1981), as well as in massive stars due to stellar
rotation (Meynet \& Maeder 2002).

The large [Mg/Fe] ratios exhibited by $\alpha$GCs thus suggest that
they could experience a more rapid star formation than that undergone
by {\sl non}-$\alpha$GCs. In turn, their larger [C/Fe] abundance
ratios might be an additional clue to constrain their formation
time-scales, although a definitive interpretation of the observed
[C/Fe] ratios is subject to the above debate on the type of stars that
preferably produce C. In fact, both scenarios --involving either
massive or intermediate-mass stars-- could qualitatively reconcile the
large [Mg/Fe] and [C/Fe] values in Figs.~\ref{index-index-fig1}$a$ and
\ref{index-index-fig3}$c$ as far as star formation in $\alpha$GCs were
brief enough to prevent Fe to incorporate into the newly formed
stars. Interestingly, the fact that $\alpha$GCs basically exhibit the
largest values of C-sensitive indices (G4300 and C$_2$4668;
Fig.~\ref{index-index-fig3}$e$) indicates that their high [C/Fe]
ratios cannot be solely driven by a lack of Fe, but a larger C
abundance is indeed necessary. The last argument, together with the
high [Mg/Fe] ratios, seem to support the hypothesis that a
non-negligible amount of the present-day C content of MR GCs in
NGC\,1407 could be supplied by massive stars during relatively rapid
star formation time-scales.

Finally, we note that some metal-poor stars in our Galaxy
also exhibit large [Mg/Fe] and [C/Fe] ratios (Kobayashi et
al.~2006). Interestingly, as reported in that paper, [Zn/Fe] may be
the clue to constrain the mechanisms driving the above trend:
hypernovae, SNe II, or even external enrichment from a binary
companion (Suda et al.~2004).

\subsubsection{The puzzle of C and N in GCs}

During the last two decades, our understanding of C and N in GCs has
basically relied on the study of the cyanogen molecule. From the
pioneering work by Burstein et al.~(1984), it was established that the
CN feature at $\sim 4215$\,\AA\ was enhanced in M31 GCs with respect
to both GGCs and E galaxies. Similar results were reported by Brodie
\& Huchra (1991) and Ponder et al.~(1998) using the CN band at $\sim
3880$\,\AA. Intriguingly, Beasley et al.~(2004; see also Trager 2003)
did not detect systematic differences between the CN$_{2}$ indices of
M31 GCs and GGCs at any metallicity, although they did still find
evidence for CN$_{2}$ enhancement in M31 GCs and GGCs with respect to
their respective galactic bulges and the family of E galaxies. To
date, such a CN {\sl anomaly} (see Trager 2003) has been reported to
exist not only for M31 GCs but also for GCs in NGC\,3115 (Kuntschner
et al.~2002), NGC\,3610 (Strader et al.~2003a) and the Fornax dSph
(Strader et al.~2003b). It is clearly inferred from Section~\ref{CN}
that NGC\,1407 is another case of such an increasingly common
phenomenon.

In order to understand the ultimate culprit of CN enhancement, there
has been an special effort to disentangle how C and N behave
separately. Making use of the NH feature at $\sim 3360$\,\AA, N
enhancement has been reported to exist for M31 GCs and GGCs (Ponder et
al.~1998; Li \& Burstein 2003), leading to the idea that N is mostly
responsible for CN variations (Burstein et al.~2004). On the other
hand, C is thought to be preferably absorbed into the CO molecule. In
fact, Li \& Burstein also present evidence for M31 and MW GCs having
G4300 indices (CH) similar to Galactic dwarf stars, which suggests
that C and N indeed have different formation mechanisms, at least in
GCs.

The puzzle of chemical evolution is even more complicated if we take
into account that C and N abundances may be linked as a consequence of
the CNO cycle. The main nucleosynthesis results concerning low and
intermediate-mass stars suggest that C yields decrease with increasing
initial metallicity, whereas N yields slightly increase with
metallicity, as N is formed by consumption of C and O already present
in the stars (Chiappini et al.~2003). If, as stated for M31 GCs, CN
enhancements are generically driven by a high N content, the striking
CN enhancement in NGC\,1407 GCs would indeed be indicative of a N
overabundance.

Under this assumption, the above picture that C and N follow opposite
trends with metallicity is, at least qualitatively, a reasonable
scenario to explain the behavior exhibited by NGC\,1407 GCs in the
CN$_2$--C$_2$4668 diagram of Fig.~\ref{index-index-fig3}$e$ (that is,
a N--C plane). Compared to MP GCs, MR GCs exhibit a striking N
overabundance at the time that C has basically saturated. In other
words, the N--C relationship steepens drastically with the increasing
metallicity as predicted by the above nucleosynthesis theories. In
fact, given that N preferably tends to form CN, C underabundances
inferred for {\sl non}-$\alpha$GCs from Fig.~\ref{index-index-fig3}$c$
could be just an artifact of the existing N overabundance: if most C
is consumed to form CN, there should not be much C left to form other
carbon molecules like C$_2$. This could be an additional reason for
the C saturation exhibited, in general, by MR GCs. Only those MR GCs
with intrinsic, high C abundances --that is, $\alpha$GCs-- possess
enough C to increase the strength of their C$_2$4668 indices
(Fig.~\ref{index-index-fig3}$f$). Furthermore, this could also explain
the fact that whilst {\sl non}-$\alpha$GCs follow the MP GC sequence
in Fig.~\ref{index-index-fig1}$a$ (with $\alpha$GCs deviating above),
$\alpha$GCs follow the MP GC sequence in
Fig.~\ref{index-index-fig3}$c$ (with {\sl non}-$\alpha$GCs deviating
below). Even though both diagrams might be essentially the same, MR
GCs shift down as a whole in Fig.~\ref{index-index-fig3}$c$ as a
consequence of the increasing efficiency of the CNO cycle with
metallicity. In any case, it seems clear that
Fig.~\ref{index-index-fig3}$f$ supports the idea that the mechanisms
dominating the formation of C and/or N have been different for MP and
MR GCs. The wide variety of CN$_2$/C$_2$4668 ratios in MR GCs
contrasts with the tight CN$_2$--C$_2$4668 sequence followed by MP
GCs, for which N and C abundances seem to increase in a parallel way
as if they both were tied to the same enrichment mechanism. Actually,
the last argument is true not only for the CN$_2$--C$_2$4668
relationship but also for other index-index diagrams involving
different element dependencies, in which MP GCs follow a tighter
abundance sequence than MR GCs do.

Interestingly, the observed behavior in NGC\,1407 GCs is in
qualitative agreement with detailed studies of C and N abundances in
individual stars of GGCs. Briley et al.~(2004) found that, apart from
a strongly anticorrelated relationship between the C depletion and the
N enhancement of main sequence stars in the MR GGC 47 Tuc (in turn,
similar to the one existing for MR GGC M71 stars), they exhibit C and
N abundances similar to those in evolved red giants, thus implying
little change in surface abundances. The last result is interpreted as
evidence for pollution or accretion events early in the history of the
cluster. The existence of bimodality in the CN content of 47 Tuc main
sequence stars (CN-strong and CN-weak; Hesser 1978) is still a
controversial issue that may shed light on the processes involved in
the formation and evolution of C and N in MR GCs. In any case, the
similarity among the C and N abundances of dwarfs and giants in MR
GGCs contrasts with the well-known relationship between the C
depletion and the luminosity of the giant branch of MP GGCs (Suntzeff
1981, Smith et al.~2005), in the sense that the larger the luminosity
the larger the C depletion and the N enhancement. In this case, the
progressive dredge-up of CNO-processed material from regions close to
the H-burning shell within Population II giants is proposed to explain
the observed trend.

It seems therefore clear that MP and MR GCs suffer intrinsically
different processes that determine their overall C and N abundances,
what is in perfect agreement with our findings for NGC\,1407 GCs. From
the point of view of C and N abundances, a general picture of GC
chemical enrichment may be summarized as follows: MP GCs form from a
relatively metal-free ISM with no apparent C depletion, as the CNO
cycle requires the prior existence of C for it to be effective. The
CNO cycle would be carried out during their evolution, thus eventually
affecting the C and N of some giant stars and, even more importantly,
polluting the ISM with enhanced N and depleted C for subsequent star
formation. MR GCs would form from such an enriched ISM that keeps
fingerprints of previous CNO processes. Pollution during the early
star formation history of the cluster could indeed contribute to
emphasize the C--N anticorrelation as well as to increase the variety
of [N/C] values among MR GCs. At this point, it is still necessary to
understand how MP GCs reach such a striking N enhancement if CNO
processes are not effective. Very massive, rotating stars (Meynet \&
Maeder 2002) or zero-metallicity, 200 -- 500\,\msun\ stars that go
hypernovae and expel out huge amounts of N, C and other elements in an
early Universe (Li \& Burstein 2003) may help to explain the existence
of primordial N overabundances.  Assuming that the nucleosynthesis
within these massive, metal-free stars should not differ much from one
to the other, these scenarios could explain not only the existence of
early N overabundances but also the fact that N and C in MP GCs follow
a well defined sequence. In other words, element enrichment in MP GCs
might be driven by a more standard mechanism whereas MR GCs may have
undergone a wider variety of enrichment histories.

\subsubsection{Different subpopulations of MR GCs?}

One of the most interesting differences among GCs is the relative
abundances of Mg and C. Could it be that such a dichotomy in the
abundance pattern is a consequence of having undergone different star
formation histories? If so, to what extent can they provide us clues
to confirm or rule out different GC formation scenarios like, e.g.,
accretions or mergers?  In order to discuss the above questions, we
examine whether the MR GCs exhibit differences in their kinematics
and/or structural parameters that may support the hypothesis of a
distinct origin.

\vspace{2.mm} i) {\it Kinematics}. The existence of asymmetries in the
velocity distribution of GC subpopulations, as well as certain
rotation patterns, can be a fingerprint of recent GC stripping,
accretion or merging scenarios (see e.g.~Bekki et al.~2005; Schubert
et al.~2007, in preparation). As mentioned in Section~\ref{radvel}, no
clear rotation has been found for our MR GCs, not even when both MR GC
sets are considered separately. In addition, we do not detect RV
asymmetries either: the mean, heliocentric RV of $\alpha$GCs is
1799\,km\,s$^{-1}$ (with a {\sl rms} standard deviation of $351$
\,km\,s$^{-1}$), whilst the one for {\sl non}-$\alpha$GCs is
1815\,km\,s$^{-1}$ (with a {\sl rms} standard deviation of $372$
\,km\,s$^{-1}$). Both values are in perfect agreement with each other,
as well as with the heliocentric RV determined for NGC\,1407 in
Section~\ref{radvel}.

\vspace{2.mm} ii) {\it Sizes}. GC sizes are thought to be determined
by both the physical conditions of the proto-cluster cloud and
subsequent dynamical evolution effects. Nevertheless, as recently
stated by H06, some of the GCs belonging to the bright-end, unimodal
population in NGC\,1407 are analogues of the high-mass GCs in the MW
(e.g.~$\omega$ Centauri and M54; Harris 1996) and M31 (G1; Meylan et
al.~2001). Ultra-Compact Dwarfs and remnant nuclei of accreted and
disrupted dEs are possible candidates to populate this bright GC
regime. Thus, could it be that either of the two MR GC subpopulations
has an {\it extragalactic} origin?

In order to test the above hypothesis, we use half-light radii for the
GCs from F06. We did not find a significant difference between the
mean sizes of $\alpha$GCs ($3.6 \pm 0.9$\,pc) and {\sl
non}-$\alpha$GCs ($3.7 \pm 1.7$\,pc). Although this result could be an
artifact of the limited spectroscopic sample we are working with, it
is important to note that our data still allows us to reproduce the
typical, mean difference of about 20 percent between MR and MP
half-light radii (Kundu \& Whitmore 1998; Larsen et al.~2001, Kundu \&
Whitmore 2001), as already reported in F06 for NGC\,1407 GCs.

\vspace{2.mm} iii) {\it Radial distribution}. The surface density
profile of MR GCs is steeper than that of MP ones (see review by
Brodie \& Strader 2006). In other words, the ratio between the number
of red and blue GC increases towards the center of the galaxy. Other
than this observational feature concerning GC spatial distribution, no
other systematic differences between GC subpopulations in Es have been
reported so far, in spite of some theoretical simulations of MR GC
formation (e.g.~through disk-disk major mergers; Bekki et al.~2002)
predict the existence of certain structures for the newly formed MR
GCs.

Regarding the projected, radial distance to the center of the galaxy
($r_{\rm p}$), we do not find any hint for $\alpha$GCs systematically
populating different regions than {\sl non}-$\alpha$GCs. Their mean
$r_{\rm p}$ values are $8.5 \pm 2.6$\,kpc and $7.0 \pm 1.2$\,kpc
respectively, with errors accounting for the {\sl rms} standard
deviation of the data. Once again, even though this analysis may be
limited by the small number of GCs, it is important to note that we do
still find significant differences between the mean $r_{\rm p}$ of MP
($11.6 \pm 1.9$\,kpc) and MR GCs ($7.8 \pm 1.4$\,kpc) as a whole.

\vspace{2.mm} To summarize, we do not find any differences between
$\alpha$GCs and {\sl non}-$\alpha$GCs in regard to kinematics, sizes
or radial distributions.  It seems therefore that, whatever the
mechanism driving the different abundance pattern is, it is not due to
distinct formation conditions as might be expected in mergers or
accretions.

\section{Summary and conclusions}
\label{summary}

We present Keck spectroscopic data for a sample of 20 GCs in
NGC\,1407, a giant E0 dominating the NGC\,1407 group of
galaxies. Radial velocities (RVs) and a subset of 20 Lick/IDS,
line-strength indices have been measured for both the GCs and the
$R_{\rm eff}/8$ central spectrum of the galaxy. On the basis of these
measurements, we have carried out an analysis of the kinematics and
stellar population properties (age, overall metallicity and certain
element abundance pattern) of both the GCs and the galaxy. Main
results are summarized in the following sections.

\subsection{Kinematics}

i) We determine a mean, heliocentric, radial velocity of $1769 \pm
70$\,km\,s$^{-1}$ --with an overall {\sl rms} standard deviation of
313\,km\,s$^{-1}$-- for the GC subsystem of NGC\,1407. No signature of
GC rotation is found to be significant, although a larger sample of
GCs is demanded to confirm this result. The derived heliocentric
radial velocity and central velocity dispersion corresponding to the
R$_{\rm eff}$/8, integrated, central spectrum of NGC\,1407 are,
respectively, $1785.6 \pm 5.2$\,km\,s$^{-1}$ and $293.5 \pm
4.1$\,km\,s$^{-1}$.

ii) The dispersion of the GC RVs clearly drops with the increasing
projected distance. This fact may have important implications on the
kind of orbits that the GCs follow within the potential well of the
galaxy. In particular, the existence of radial orbits (rather than
isotropic orbits) would help to explain the observed trend.

\subsection{Age, metallicity and element abundance ratios}

Making use of the measured Lick indices and the SSP model predictions
by TMB03 and TMK04, ages, metallicities and [$\alpha$/Fe] ratios for
the GCs and the R$_{\rm eff}$/8, integrated, central spectrum of the
galaxy have been derived using two different approaches: i) the
$\chi^2$ procedure by Proctor et al.~(2004), and ii) an iterative
procedure based on Mg, Fe and Balmer Lick indices (see details in
Section~\ref{iterative}). An additional metallicity estimate for the
GCs has been obtained using the PCA method of Strader \& Brodie
(2004). On the basis of the above methods we conclude that:

\subsubsection{Metallicities}

iii) As expected for a massive E, the GC metallicity distribution
ranges from metal-poor to slightly above solar. MP and MR GC
subpopulations peak at [Z/H] $\sim -0.9$ and $\sim -0.2$\,dex with
{\sl rms} standard deviations of 0.28 and 0.16\,dex respectively. The
integrated, $R_{\rm eff}$/8 central spectrum of NGC\,1407 is
metal-rich ([Z/H] $\sim +0.35$\,dex).

\subsubsection{Ages}

iv) Most GCs in NGC\,1407 are confirmed to be old. Depending on the
age-dating method, we obtain mean values of $10.9 \pm 0.5$\,Gyr (with
a {\sl rms} standard deviation of 2.1\,Gyr) and $11.7 \pm 0.4$\,Gyr
(with a {\sl rms} standard deviation of $\sim 1.6$\,Gyr). Analogously,
the integrated, $R_{\rm eff}$/8 central spectrum of NGC\,1407 is also
found to be similarly old ($11.9 \pm 1.4$\,Gyr or $12.0 \pm
0.4$\,Gyr).

v) In spite of most GCs being old, both age-dating techniques
consistently find evidence for the existence of three, presumably {\sl
young} GCs, with mean ages of $\sim 4$\,Gyr. However the BHB spectral
diagnostic proposed by SCH04 indicates that these presumably {\it
young} GCs are, actually, old GCs hosting BHB stars.

\subsubsection{Abundance ratios}

vi) GCs in NGC\,1407 exhibit an evident overabundance of
$\alpha$-elements. We find them to have mean [$\alpha$/Fe] ratios of
$\sim +0.31$\,dex, although it quite depends on the procedure carried
out ($+0.24 \pm 0.03$\,dex, with a {\sl rms} standard deviation of
0.11\,dex, or $0.38 \pm 0.04$\,dex, with a {\sl rms} standard
deviation of 0.18\,dex). For the integrated, $R_{\rm eff}$/8 central
spectrum of NGC\,1407 we derive similar [$\alpha$/Fe] values of $\sim
+0.33$\,dex ($+0.33 \pm 0.01$ and $+0.34 \pm 0.01$\,dex from each
different technique).

vii) Concerning single element abundances, we report --for the first
time-- hints for the existence of two chemically-distinct
subpopulations of metal-rich (MR) GCs. Some MR GCs exhibit
significantly larger [Mg/Fe] {\it and} [C/Fe] ratios as compared to
the rest of MR GCs. This is the first evidence for a correlation
between Mg and C abundances among extragalactic GCs. The possibility
that high [Mg/Fe] and [C/Fe] ratios could be just driven by a Fe
depletion has been ruled out, since GCs exhibiting high [C/Fe] ratios
indeed populate the high-metallicity regime of C-sensitive index-index
diagrams. A true enhancement of their C and Mg content therefore is
favored. Different star-formation time-scales have been proposed to
explain the intriguing correlation between Mg and C abundances. If
this were the case, a non-negligible amount of the present-day C
content of MR GCs in NGC\,1407 could be supplied by massive stars
during relatively short star formation time-scales.

viii) We also find striking CN overabundances over the full GC
metallicity range. Interestingly, the behavior of C and N in MP GCs
clearly deviates from the one exhibited by MR GCs, qualitatively
resembling the existing differences among MP and MR GGCs. In
particular, for MR GCs, N increases drastically at the same time that C
essentially saturates and becomes clearly depleted as compared to the
model predictions. From the point of view of chemical evolution, this
may be interpreted as a consequence of the increasing importance of
the CNO cycle with increasing metallicity.

ix) The above observational constrains allow us to propose the
following scenario for GC chemical enrichment. Assuming a primordial
overabundance of N, MP GCs would form from a relatively metal-free ISM
with no apparent C depletion, as the CNO cycle requires the prior
existence of C for it to be effective. The CNO cycle would be carried
out during their evolution, thus eventually affecting the C and N of
some giant stars and, more importantly, polluting the ISM with
enhanced N and depleted C for subsequent star formation. MR GCs would
form from such an enriched ISM that preserves the previous CNO
products. Pollution during the early star formation history of the
cluster could indeed contribute to the C--N anticorrelation as well as
increasing the variety of [N/C] values among MR GCs.  Overall, the
data presented in this paper support the idea that element enrichment
in MP GCs might be driven by a more standard mechanism whereas MR GCs
may have undergone a wider variety of enrichment histories.

\subsection{GC formation scenarios}

x) The fact that the two chemically-distinct subpopulations of MR GCs
do not exhibit significant differences in their radial velocities,
sizes and projected galactocentric distances is at odds with the
hypothesis that any of the two subpopulations could have an {\sl
extragalactic} origin such as recent accretion or merger
processes. Although we strictly cannot rule out the last possibility,
local, intrinsic differences in the chemical composition of protoGCs
haloes within the potential well of the main galaxy halo (e.g.~Forbes
et al.~1997), together with different star formation time-scales, may
suffice to account for the observed dichotomy.

xi) Overall, the old ages inferred for the GCs in NGC\,1407 are
consistent with the idea that both MP and MR GC subpopulations in
massive Es were formed at high redshift (Strader et
al.~2005). Unfortunately, whether or not the MR subpopulation is
slightly (say $2 - 3$\,Gyr) younger than the MP GCs is still an open
question. The difficulty of detecting small age differences among old
GC subpopulations has been illustrated in this paper, as BHB effects,
different age-dating techniques, systematics of the models, and
typical uncertainties in the data may blur the derived results. Larger
samples of high-quality GC spectra are strongly demanded to address,
from a reliable, statistical point of view, the epochs of GC formation
in the Universe.

\section*{Acknowledgments}

The authors are indebted to the anonymous referee for very useful
comments and suggestions. Gratitude is also due to Juerg Diemand for
his work on preliminary cosmological simulations for our GC set, as
well as to Lee Spitler for providing us with his GC size
measurements. This work has been mainly supported by NSF grants AST
02-06139 and AST 05-07729, and the Spanish research projects
AYA2003-01840 and AYA2004-03059. AJC acknowledges financial support
from a Spanish Fundaci\'on del Amo Fellowship. AJC is a {\sl Juan de
la Cierva} Fellow of the Spanish Ministry of Education and
Science. The authors wish to recognize and acknowledge the very
significant cultural role and reverence that the summit of Mauna Kea
has always had within the indigenous Hawaiian community. We are most
fortunate to have the opportunity to conduct observations from this
mountain.

\clearpage

\begin{figure*}[!t]
\center{
\scalebox{0.74}{\includegraphics[angle=-90]{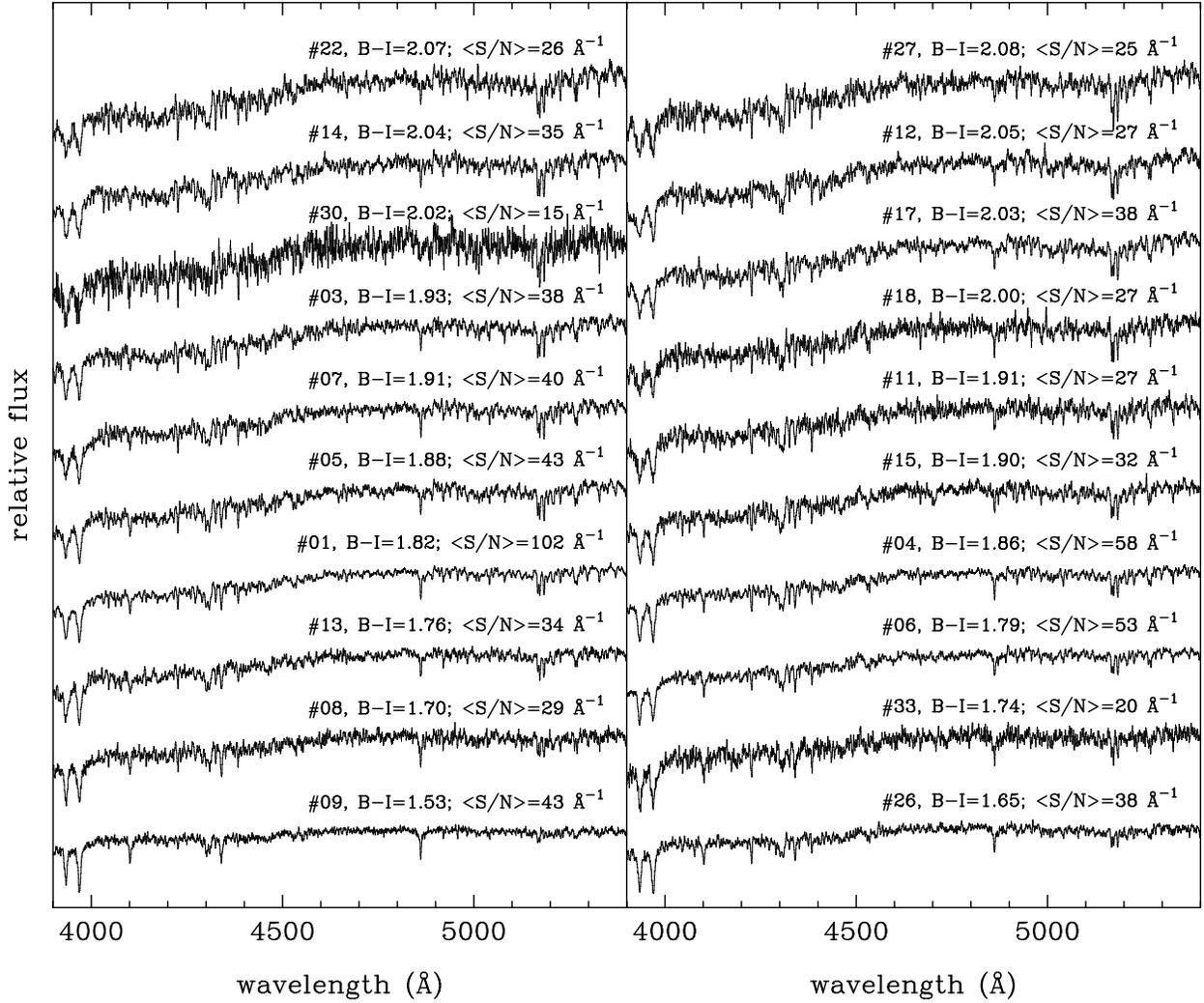}}
\caption{Spectra of the 20 GCs in NGC\,1407 observed in this
work. From bottom to top and from right to left, the GC spectra are
displayed following an increasing color sequence. ID numbers, $B - I$
colors, and mean signal-to-noise ratios per angstrom averaged over the
whole spectral range are provided in the labels.}
\label{GCspec}
}
\end{figure*}

\begin{figure*}[!t]
\center{
\scalebox{0.5}{\includegraphics[angle=-90]{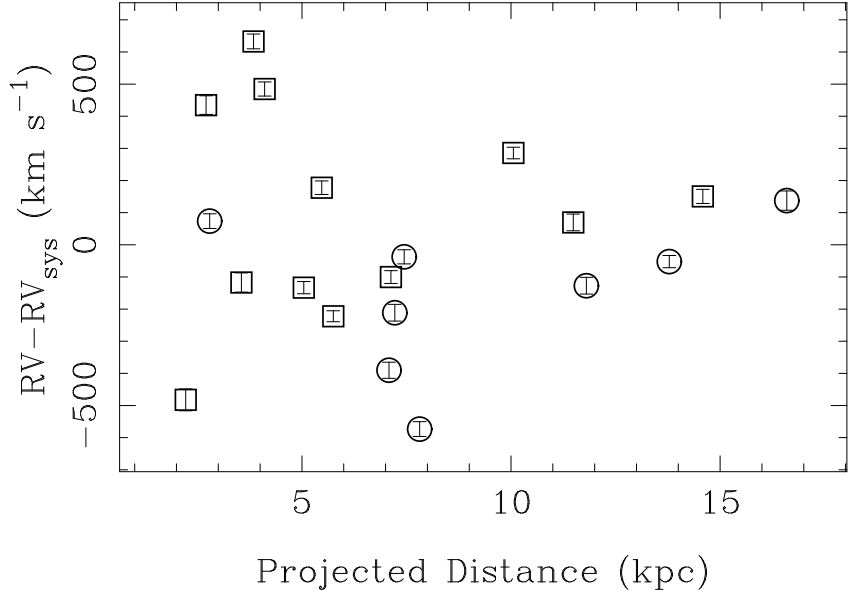}}
\caption{Radial velocities of the GCs corrected for the systemic
radial velocity (RV$_{\rm sys} = 1769 \pm 70$\,km\,s$^{-1}$) as a
function of the projected distance from the center of
NGC\,1407. Different symbols represent GCs with $B-I < 1.87$ (circles;
metal-poor) and $B-I > 1.87$ (squares; metal-rich). A distance of
20.9\,Mpc has been assumed for the galaxy (Forbes et al.~2006).}
\label{RV-r}
}
\end{figure*}

\begin{figure*}[!t]
\center{
\scalebox{0.7}{\includegraphics[angle=-90]{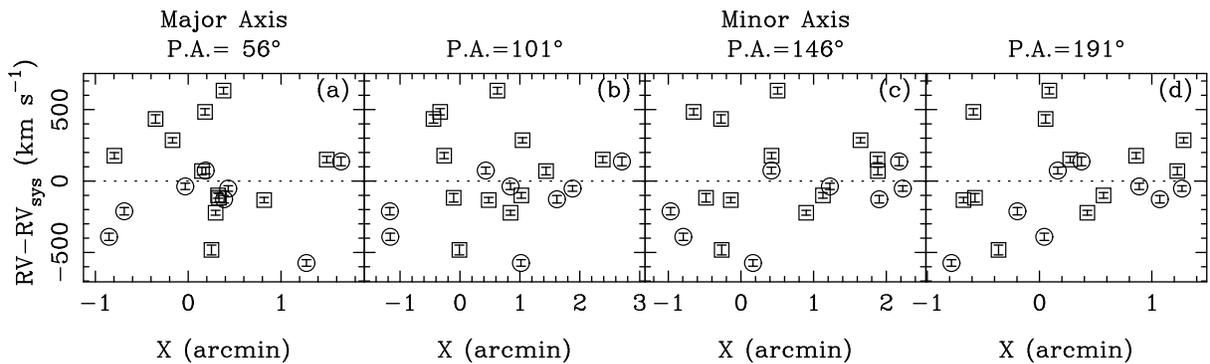}}
\caption{Relative radial velocities of the GCs as a function of
projected distance to different axis: major axis ($a$), minor axis
($c$) and two intermediate directions ($b$ and $d$). No systematic
rotation of the GC system around the galaxy is detected. Symbols as in
Fig.~\ref{RV-r}.}
\label{checkrot}
}
\end{figure*}

\begin{figure*}[!t]
\center{
\scalebox{0.70}{\includegraphics[angle=-90]{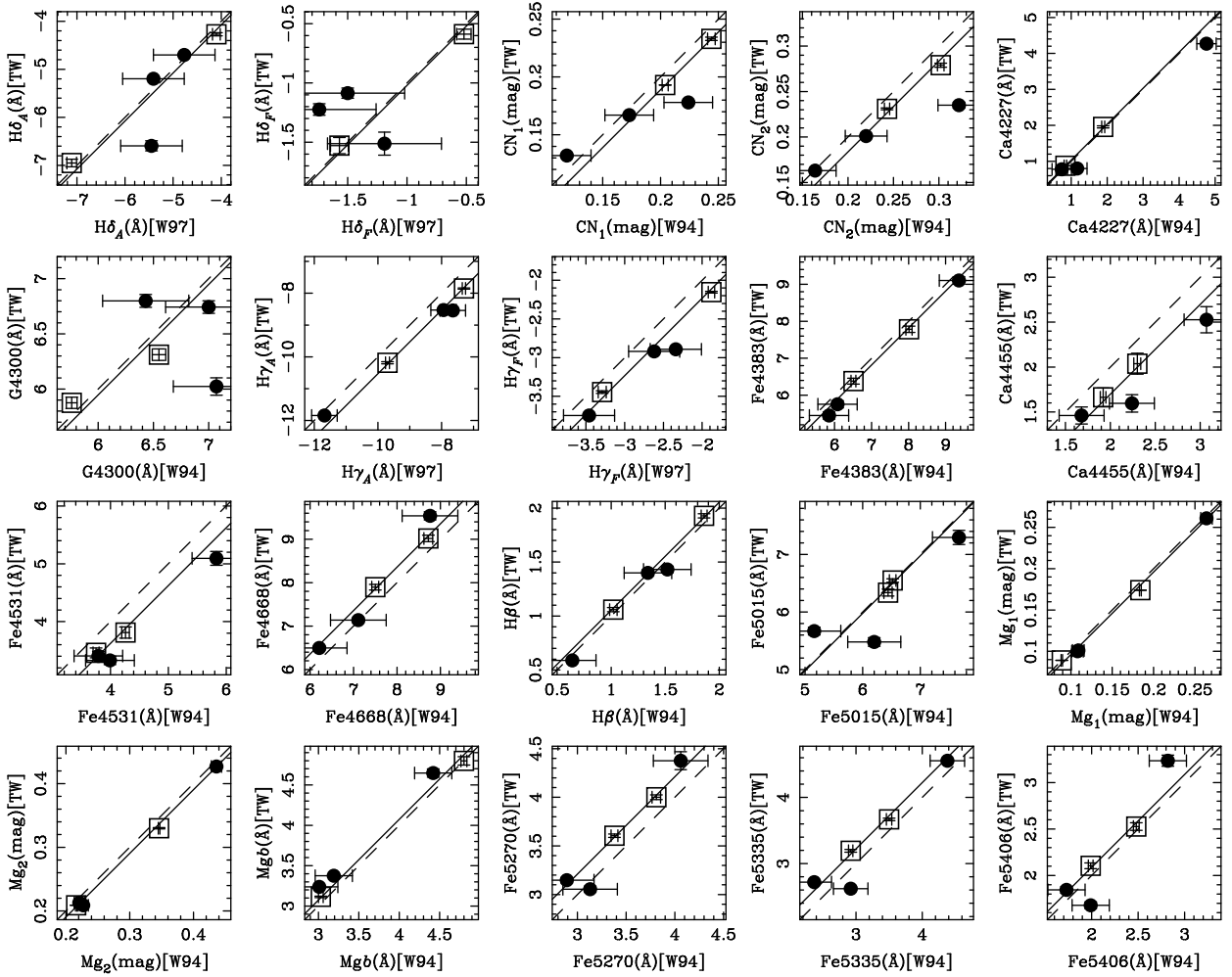}}
\caption{Comparison between the Lick indices of five Lick/IDS stars in
the Lick system (W94; W97) and those measured in this work (TW) at the
spectral resolutions of Table~\ref{Lickset}. The mean index errors
given in W94 and W97 are assumed for HR\,2153, HR\,2459, and HR\,8165
(filled circles). For the Lick  stars HR\,2002 and
HR\,7429 (open squares), which have several repeat observations in W94
and W97, their typical errors are used. 
Dashed lines indicate the 1:1 relation, while solid lines
represent the derived, error-weighted offsets between both
datasets. }}
\label{lickoffs}
\end{figure*}

\begin{figure*}[!t]
\centering{
\scalebox{0.40}{\includegraphics[angle=-90]{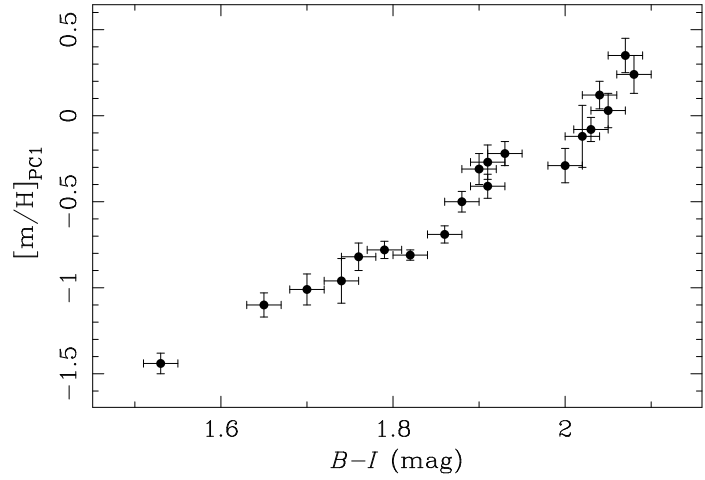}}
\caption{Color--metallicity relationship for GCs in
NGC\,1407. Metallicities are derived from the first component (PC1) of
the principal component analysis (PCA) for GGCs by Strader \&
Brodie~(2004). $B-I$ colors are taken from Forbes et al.~(2006; see
also Table~\ref{basicdata}).}
\label{PCA-color}
}
\end{figure*}

\begin{figure*}[!t]
\centering{ \scalebox{0.50}{\includegraphics[angle=0]{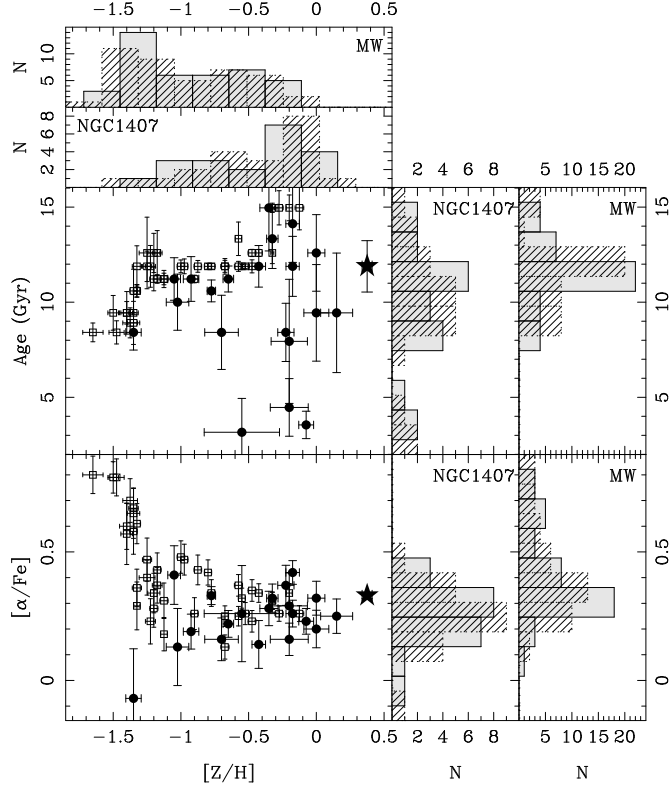}}
\caption{Ages, metallicities and [$\alpha$/Fe] ratios for the GCs in
NGC\,1407 (filled circles) and the central ($R_{\rm eff}$/8),
integrated, galaxy spectrum (filled star) as derived from the $\chi^2$
procedure by Proctor et al.~(2004; see text). Open squares represent
the sample of 41 GGCs by Schiavon et al.~(2005) which has been
analyzed in the same way for the sake of comparison. Histograms of the
derives ages, metallicities and [$\alpha$/Fe] ratios are provided for
both samples of GCs. In each case, shaded and stripped histograms
--shifted half a bin one from the other-- account for bin sampling
effects.}
\label{fighistog}
}
\end{figure*}

\begin{figure*}
\center{
\scalebox{0.70}{\includegraphics[angle=0]{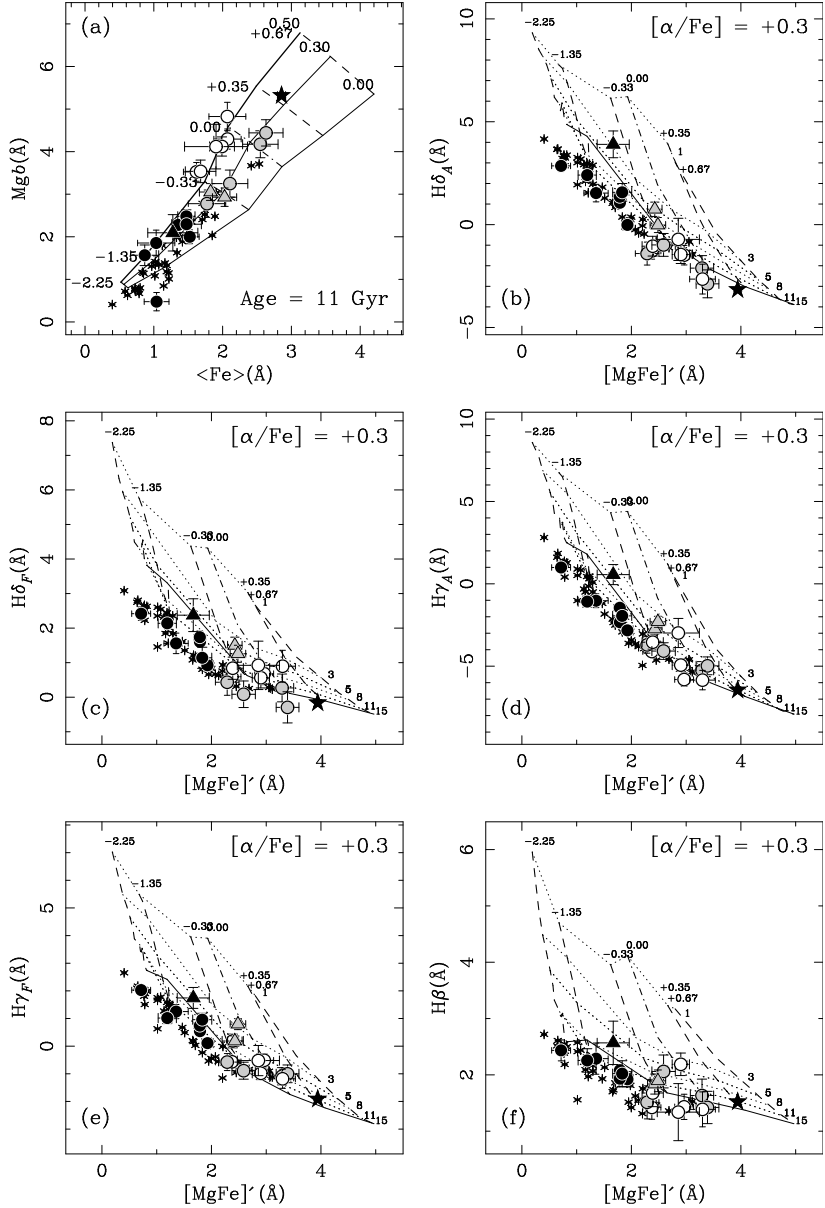}}
\caption{\small Age-metallicity-[$\alpha$/Fe] diagnostic plots constructed on
the basis of Mg$b$, $<$Fe$>$, [MgFe]' and the Balmer Lick
indices. Circles (both filled and open) and triangles correspond,
respectively, to old and {\it young} GCs ({\sl y}GCs) as derived from
the $\chi^2$ procedure in Section~\ref{chi2}. Metal-poor GCs ($B - I <
1.87$) are illustrated as black-filled symbols, whilst open and
grey-filled symbols represent metal-rich GCs ($B - I > 1.87$).  The
star corresponds to the $R_{\rm eff}/8$ central spectrum of
NGC\,1407. Asterisks represent the sample of 41 GGCs of Schiavon et
al.~(2005). In panel $a$, model predictions for Mg$b$ and $<$Fe$>$ at
fixed age (11\,Gyr) and varying metallicity and [$\alpha$/Fe] ratios
(dashed lines and different-width solid lines respectively; see the
labels) are over-plotted. In particular, metal-rich GCs with
[$\alpha$/Fe]$ > 0.4$\,dex --as derived from this plane-- are
displayed as open circles ($\alpha$GCs), whilst grey-filled symbols
are reserved for the rest of metal-rich GCs ({\sl
non}-$\alpha$GCs). Panels $b - f$ display the model predictions with
constant [$\alpha$/Fe] ratio ($+0.3$) and varying age (dotted lines;
see the labels) and metallicity for the Balmer Lick indices and
[MgFe]'. To guide the eye, the 15\,Gyr line is represented as a solid
line and the solar-metallicity one as dash-dotted line.}
\label{index-index-fig1}
}
\end{figure*}

\begin{figure*}[!t]
\center{
\scalebox{0.80}{\includegraphics[angle=0]{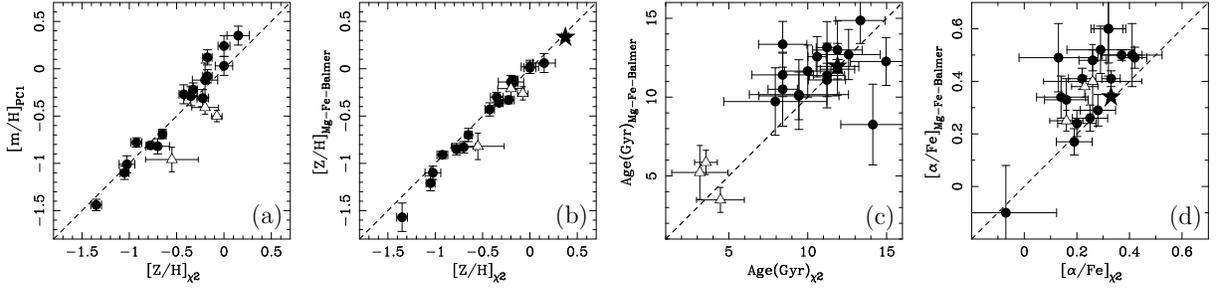}}
\caption{Comparison between the ages, metallicities and [$\alpha$/Fe]
ratios of the GCs derived from the PCA calibration (Strader \& Brodie
2004), the $\chi^2$ procedure (Proctor et al.~2004) and the iterative
procedure based on Mg, Fe and Balmer Lick indices (see details in
Section~\ref{iterative}). Open triangles have been used for those
presumably young GCs, whereas filled circles represent
the remaining of GCs. The central ($R_{\rm eff}$/8), integrated, galaxy
spectrum is represented as a filled star. The dashed line indicates
the 1:1 relationship.}
\label{figcomp4}
}
\end{figure*}

\begin{figure*}[!t]
\center{
\scalebox{0.70}{\includegraphics[angle=0]{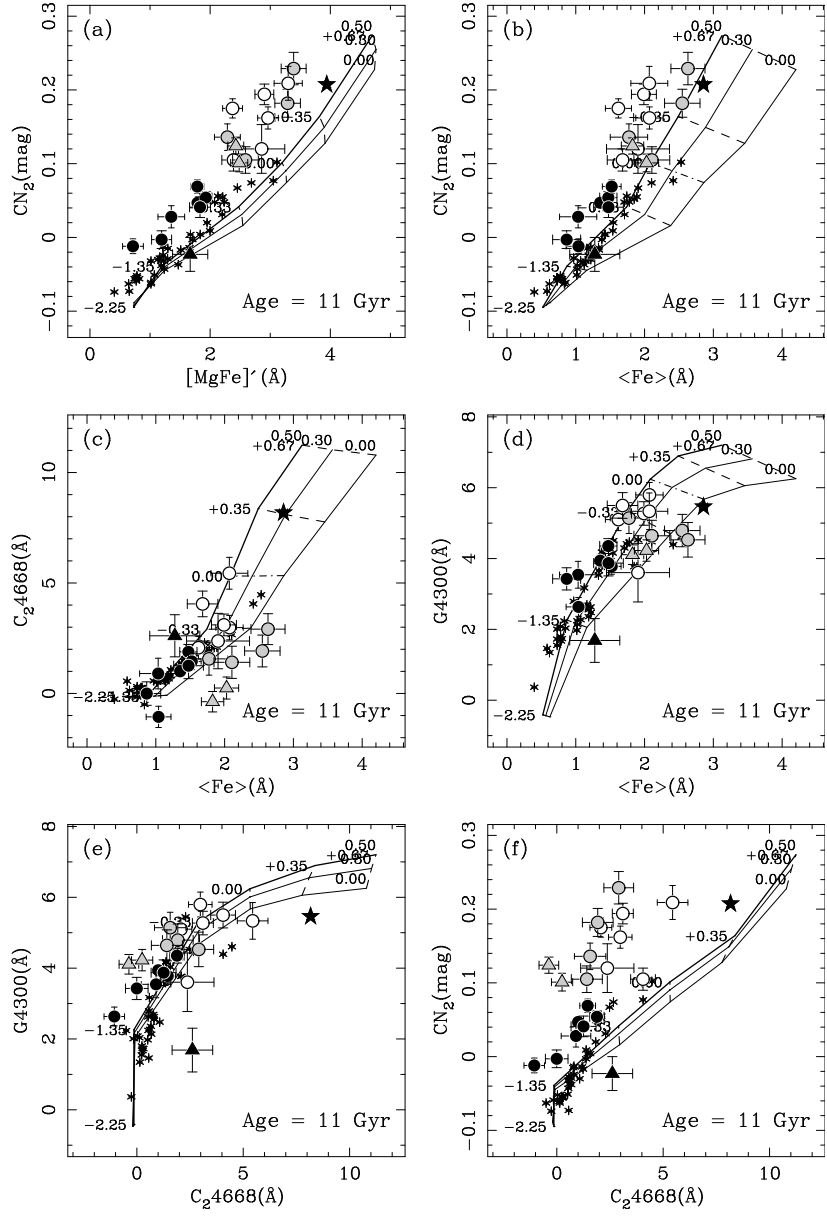}}
\caption{Index-index diagnostic diagrams of Lick indices
sensitive to C, N, Mg and Fe. Symbols and models as in
Fig.~\ref{index-index-fig1}}
\label{index-index-fig3}
}
\end{figure*}

\begin{figure*}[!t]
\center{
\scalebox{0.4}{\includegraphics[angle=-90]{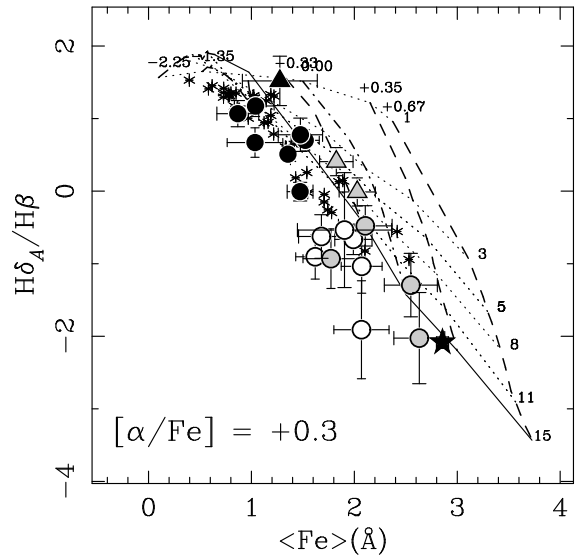}}
\caption{Diagnostic diagrams to detect BHB morphologies in the
integrated spectra of old GCs as proposed by Schiavon et
al.~(2004). Symbols and models as in Figure~\ref{index-index-fig1}.}
\label{figBHB}
}
\end{figure*}

\end{document}